\documentclass[dvipsnames]{article}

\usepackage[normalem]{ulem}
\usepackage{a4wide}

\usepackage{amsmath,amsfonts,amsthm,amssymb,bbm}
\usepackage[justification=justified, singlelinecheck=false]{caption}
\usepackage{graphicx}
\usepackage{enumitem}

\usepackage{tikz}
\usepackage{algorithmicx}
\usepackage{algorithm}
\usepackage{algpseudocode}

\usepackage[scaled=.90]{helvet}
\usepackage{courier}
\usepackage{ae}
\usepackage[T1]{fontenc}
\usepackage{subfigure}
\usepackage{graphicx,color}
\usepackage{xcolor}

\usepackage[T1]{fontenc}
\usepackage{here} 
\usepackage[colorlinks=false, pdfborder={0 0 0}]{hyperref}
\usepackage{comment}
\usepackage{array}
\usepackage{xpatch}
\makeatletter
\xpatchcmd{\@thm}{\thm@headpunct{.}}{\thm@headpunct{}}{}{}
\makeatother

\usepackage[title]{appendix} 

\newtheorem{remark}{Remark}

\theoremstyle{definition}

\begin{document}

\title{Approximate Bayesian computation for stochastic hybrid systems with empirically observed ergodic behaviour}

\author{Sascha Desmettre\footnotemark[1]\thanks{Institute of Financial Mathematics and Applied Number Theory, Johannes Kepler University Linz (sascha.desmettre@jku.at)}, Agnes Mallinger\footnotemark[2]\thanks{Institute of Stochastics, Johannes Kepler University Linz (agnes.mallinger@jku.at)}, Amira Meddah\footnotemark[3]\thanks{Institute of Stochastics, Johannes Kepler University Linz (amira.meddah@jku.at)}, Irene Tubikanec\footnotemark[4]\thanks{Institute of Applied Statistics, Johannes Kepler University Linz (irene.tubikanec@jku.at)}\\}
\date{}
\maketitle		

\thispagestyle{empty}

\section*{Abstract}

Piecewise diffusion Markov processes (PDifMPs) form a versatile class of stochastic hybrid systems that combine continuous diffusion processes with discrete event-driven dynamics, enabling flexible modelling of complex real-world hybrid phenomena. The practical utility of PDifMP models, however, depends critically on accurate estimation of their underlying parameters. In this work, we present a novel framework for parameter inference in PDifMPs based on approximate Bayesian computation (ABC). Our contributions are threefold. First, we provide detailed simulation algorithms for PDifMP sample paths. Second, we extend existing ABC summary statistics for diffusion processes to account for the hybrid nature of PDifMPs, showing particular effectiveness for ergodic systems. Third, we demonstrate our approach on several representative example PDifMPs that empirically exhibit ergodic behaviour. Our results show that the proposed ABC method reliably recovers model parameters across all examples, even in  challenging scenarios where only partial information on jumps and diffusion is available or when parameters appear in state-dependent jump rate functions. These findings highlight the potential of ABC as a practical tool for inference in various complex stochastic hybrid systems.

\subsubsection*{Keywords} Simulation-based inference, Approximate Bayesian computation, Parameter estimation, Stochastic hybrid systems, Piecewise diffusion Markov processes.

\subsubsection*{MSC Subject classifications} 60H35, 62F15, 93C30 

\subsubsection*{Acknowledgements} 
We thank JKU Linz, in particular the vice rectorates for Finance and Entrepreneurship and for Research and International Affairs, for supporting this work with means of their 8th Invest Call.


\section{Introduction}
\label{sec:1:Introduction}
\vspace{-0.2cm}

Combining continuous-time and event-driven dynamics, stochastic hybrid systems (SHS) have emerged as a powerful modelling framework over the past decades \cite{Bujorianu2006,cassandras2018SHS, davis1984piecewise}. They have been applied across a wide range of fields, ranging from mathematical biology \cite{berg1972chemotaxis, buckwar2023stochastic,buckwar2025numerical,cloez2017probabilistic, meddah2024stochastic} and neuroscience \cite{buckwar2011exact,pakdaman2010fluid} to biochemistry \cite{singh2010stochastic} and mathematical finance \cite{buckwar2024americanoptionpricingusing,ishijima2011regime}. 

A prominent subclass of SHS is formed by piecewise deterministic Markov processes (PDMPs), which combine deterministic continuous dynamics -- typically modelled by ordinary differential equations (ODEs) -- with stochastic jump dynamics. PDMPs were first introduced by Davis in 1984 \cite{davis1984piecewise}. A few years later, Blom \cite{blom1988piecewise} proposed a more general model in which the continuous dynamics are governed not by ODEs but by stochastic differential equations (SDEs), and state-dependent reset maps were introduced, leading to the formulation of piecewise diffusion Markov processes (PDifMPs). The theoretical foundations of these models were subsequently established by Bujorianu  and Lygeros in  2003~\cite{bujorianu2003reachability}. Notably, jump-diffusion models \cite{Runggaldier2003} form a related class of systems that also combine SDE dynamics with random jumps. However, PDifMPs provide a more flexible modelling framework by allowing for state-dependent regime changes, thereby capturing a broader range of hybrid stochastic behaviours.

From a computational perspective, simulating PDifMPs remains challenging due to the interaction between stochastic diffusion dynamics and discrete jumps. Dedicated numerical methods have recently been proposed and shown to converge to such hybrid systems \cite{buckwar2025numerical}, and exact schemes for first-passage times have been developed to capture threshold-dependent events \cite{desmettre2025hybrid,DesmettreKhuranaMeddah2025}. Despite these advances, suitable statistical inference methods for PDifMPs, which enable parameter estimation, remain largely uninvestigated in the literature.

A PDifMP may depend on several parameters that govern both its SDE and jump process dynamics. In this work, we focus on parameter estimation using approximate Bayesian computation (ABC), a simulation-based approach for Bayesian parameter inference that is particularly useful when the underlying likelihood function is unknown or intractable. Over the past decades, ABC has become one of the most widely used and well-established tools for inference in complex and realistic mathematical models (see \cite{Marin2012,sisson2018handbook} for reviews). Originally introduced in the context of population genetics \cite{Beaumont2002}, it has since been successfully applied to a wide range of problems arising in diverse application areas. More recently, ABC has been employed for parameter inference in SDE models \cite{Jovanovskietal2024,Kypraios2017,PicchiniTamborrino2022}, particularly for ergodic SDEs \cite{Buckwar2020,Ditlevsenetal2023,SAMSON2025108095}, and has also received limited attention in the context of jump-diffusion models \cite{Creel2015,Frazier2019}. To our knowledge, ABC has not yet been applied to parameter inference in PDifMPs.

Adapting the ABC method for inference in PDifMPs involves two major challenges. First, it requires efficient simulation of synthetic datasets (sample paths) of the PDifMP. We address this by proposing detailed algorithms for path simulation: one for constant jump rate functions and another for state-dependent jump rates. Second, the quality of ABC strongly depends on informative summary statistics. To this end, we extend the summary statistics proposed in \cite{Buckwar2020} for SDE models to PDifMPs, taking into account their hybrid nature. These statistics are particularly effective for PDifMPs with ergodic behaviour, as ergodicity allows relevant information to be extracted from individual datasets rather than requiring repeated simulations for each parameter~configuration.

We assess the performance of our inference method on four representative PDifMP examples that empirically exhibit ergodic behaviour. Our approach achieves strong inference results across all examples, including challenging scenarios where only the number of jumps (rather than their exact timing) is observed, the SDE component is only partially observed, and/or  parameters appear in a state-dependent jump rate function. These results demonstrate the potential of ABC, and in particular our proposed method, for inference in PDifMPs.

This paper is organised as follows. In Section \ref{sec:2:PDifMPs}, we recall the general PDifMP framework and introduce detailed algorithms for simulating PDifMP sample paths. In Section \ref{sec:3:PDifMPs}, we present several representative example PDifMPs and demonstrate empirically that they exhibit ergodic behaviour. In Section \ref{sec:4:ABC}, we review the ABC method and describe its adaptation for inference in PDifMP models with ergodic behaviour. The corresponding inference results are reported in Section~\ref{sec:5:Inference},  followed by a detailed examination of a test problem exhibiting distinct behaviour in Section~\ref{sec:6:TP3}. 
Finally, Section~\ref{sec:7:conclusion} concludes with a summary and an outlook on future research directions. Accompanying \texttt{R} code is provided on GitHub (cf. Section \ref{sec:implDet}).


\section{Piecewise diffusion Markov processes}
\label{sec:2:PDifMPs}

This section discusses PDifMPs. In Section \ref{sec2:1:GeneralFramework}, we recall the general framework of PDifMPs. In Section \ref{sec2:2:Simulation}, we describe in detail how they can be simulated. In Section \ref{sec:3:PDifMPs}, we introduce a diverse set of representative model problems from the class of PDifMPs with empirically observed ergodic behaviour, which serve as the basis for the analyses in the subsequent sections. 


\subsection{General framework of PDifMPs}
\label{sec2:1:GeneralFramework}

Let $T>0$ and $m\in \mathbb{N}$. Let $(\Omega,\mathcal{F},\mathbb{P})$ be a complete probability space with a complete and right-continuous filtration $(\mathcal{F}_t)_{t\in[0,T]}$. Moreover, let $(W_t)_{t\in [0,T]}$ be an $m$-dimensional standard Wiener process on $(\Omega,\mathcal{F},\mathbb{P})$ and adapted to $(\mathcal{F}_t)_{t\in[0,T]}$.

A PDifMP is a stochastic process $U=(U_t)_{t\in[0,T]}$ adapted to $(\mathcal{F}_t)_{t\in[0,T]}$ consisting of two possibly interacting components, i.e. $U_t:=(X_t,Z_t)$ for all $t\in [0,T]$. Its state space is given by $E= D_1 \times D_2$ with $D_1 \subseteq \mathbb{R}^{d_1}$ and $D_2 \subseteq \mathbb{R}^{d_2}$, for $d_1,d_2 \in \mathbb{N}$. The Borel $\sigma$-algebra $\mathcal{B}(E)$ is generated by the topology of $E$.

\paragraph{Description of PDifMP components}

The component $Z=(Z_t)_{t\in [0,T]}$ is a jump process with right-continuous, piecewise-constant paths and takes values in $D_2$. Specifically, the (random) jump times $(J_k)_{k\in  \{0,\ldots, N^j-1 \}}$, where $N^j \in \mathbb{N}$ is the number of jumps occurring within $[0,T]$, are governed by a rate function $\Lambda$, which may depend on both components of the PDifMP. 
On each interjump interval $[J_k,J_{k+1})$, the process $Z_t$ remains constant, i.e., $Z_t \equiv z_k$, for all $t \in [J_k,J_{k+1})$, where $z_k\in D_2$.

The component $X=(X_t)_{t\in [0,T]}$ is a diffusion process with continuous paths and takes values in $D_1$. More precisely, on each interjump interval $[J_k,J_{k+1})$, it follows the dynamics of the SDE
\begin{equation}
\begin{aligned}
        dX_t &= F(X_t,z_k)dt + \Sigma(X_t,z_k) dW_t, \quad t \in [J_k,J_{k+1}), \\
        X_{J_k} &= x_{J_k}, \quad x_{J_k} \in D_1,
\end{aligned}
\label{SDE}
\end{equation}
where the drift coefficient $F: E\to \mathbb{R}^{d_1}$ and the diffusion coefficient $\Sigma: E \to \mathbb{R}^{d_1 \times m}$ depend on the corresponding value $z_k$ of the jump process $Z_t$.

Denote by $\phi$ the solution (flow) of SDE \eqref{SDE}, i.e.
\begin{equation*}
    X_t=\phi(t-J_k,u_k), \quad t \in [J_k,J_{k+1}),
\end{equation*}
where $u_k:=(x_{J_k},z_k)$. The next initial condition $X_{J_{k+1}} = x_{J_{k+1}}$ for the SDE over the subsequent interjump interval $[J_{k+1}, J_{k+2})$ is obtained by $\phi(J_{k+1}-J_k,u_k)$. This means that although $\phi$ is defined only on the interval $[J_k,J_{k+1})$, its right-limit value at $J_{k+1}$ is used to define the next initial condition, thereby ensuring the continuity of the paths for the process $X_t$. Moreover, the next value $z_{k+1}$ of the jump process $Z_t$ is drawn according to a transition kernel $Q(x_{J_{k+1}}, z_k, \cdot)$, which may depend on both components $X_t$ and $Z_t$ of the PDifMP. 

Note that we define the initial jump time as $J_0:=0$ and the initial condition $u_0=(x_0,z_0)$ is a given value from $E$.  

\paragraph*{Characteristic triple}

A PDifMP can be defined by its characteristic triple $(\phi, \Lambda, Q)$:

\begin{itemize}
    \item[-] $\phi: [0,T] \times E \to D_1$ is the flow associated with the piecewise-defined SDE \eqref{SDE}.
    \item[-]$\Lambda: E \to \mathbb{R}_+$ is a non-negative measurable map, that defines the jump rate of the PDifMP. 
      \item[-] $Q:E\times \mathcal{B}(D_2) \to  [0,1]$ is a transition kernel that assigns to each $u=(x,z) \in E$ a probability measure $Q(u, \cdot)$ on $(D_2, \mathcal{B}(D_2))$, describing the distribution of the post-jump state of $Z$.
\end{itemize}

\noindent We refer the reader to Section \ref{sec:3:PDifMPs} for concrete choices of  $(\phi, \Lambda, Q)$ and to Remark \ref{rem:extensions} in that section for possible extensions.

\paragraph*{Assumptions}
We assume the following regularity conditions.
\begin{enumerate}
    \item  The drift $F$ and the diffusion coefficient $\Sigma$ are such that there exists a unique strong solution for SDE~\eqref{SDE}. (Standard conditions are global Lipschitz continuity and linear growth \cite{oksendal2010sde}.) 
    \item The jump rate function $\Lambda$ is such that the probability of infinitely many jumps occurring within the finite time interval $[0,T]$ is zero. 
     \item The jump rate function $\Lambda$ is such that the probability that at least one jump occurs within $[0,T]$ is positive.
\end{enumerate}


\subsection{Simulation of PDifMPs}
\label{sec2:2:Simulation}

In this section, we detail algorithms for the simulation of paths of PDifMPs. The full pseudo code can be found in Appendix \ref{appA.1}.

\paragraph{Simulation of SDE paths over interjump intervals}

A key step for the simulation of PDifMP paths is the simulation of paths of the solution of SDE \eqref{SDE} over a given sample interjump interval $[j_k,j_{k+1})$, where $j_k$ and $j_{k+1}$ are realisations of $J_k$ and $J_{k+1}$, respectively. This requires to discretise the considered time interval $[j_k,j_{k+1}]$ including the end point $j_{k+1}$ with step size $h>0$. Since the length $\tau=j_{k+1}-j_k$ might not be a multiple of $h$, the size of the last step can be smaller than $h$. Specifically, we obtain the time grid
\begin{equation*}
    t_n^k = j_{k} + n h, \quad n=0,\ldots,N_k,  \quad \text{where } N_k=\left\lfloor \frac{j_{k+1} - j_k}{h} \right\rfloor,
\end{equation*}
with final time point $t_{N_k+1}^k:={j_{k+1}}$, i.e. the size of the last step is $\tau-N_kh$ and $\lfloor \cdot \rfloor$ denotes the integer floor function.  Figure \ref{fig:adaptiveGrid} reports an example time grid for $N_k=9$. Algorithm~\ref{alg:PathFlowPhi} in Appendix~\ref{appA.1} summarises the simulation procedure of an SDE path over a general time grid.

\begin{figure}[t]
    \centering
    \includegraphics[width=\textwidth]{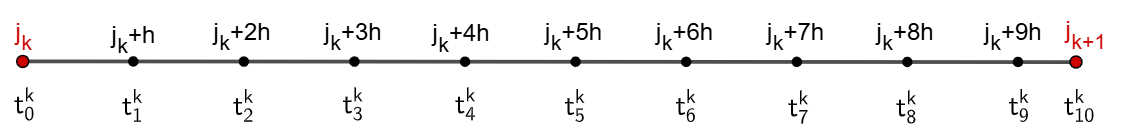}
    \caption{Example time grid for $N_k=9$.}
    \label{fig:adaptiveGrid}
\end{figure}

\paragraph{Simulation of PDifMP paths with constant jump rate}

For simplicity, we first consider a constant jump rate function $\Lambda \equiv \lambda > 0$. The construction of a PDifMP path then works according to the following steps.

Fix the initial jump time $j_0=0$ and the initial condition $u_0=(x_0,z_0) \in E$, and set $k=0$. Simulate the next jump time by generating a waiting time $\tau$ from the exponential distribution with rate $\lambda$, and set $j_{k+1}=j_{k}+\tau$.  Given $u_k=(x_{j_k},z_k)$, generate a path of the SDE solution
\begin{equation*}
    X_t=\phi(t-j_k,u_k), \quad t \in [j_k,j_{k+1}],
\end{equation*}
cf. Algorithm \ref{alg:PathFlowPhi}. Then, draw the the next constant $z_{k+1}$ from the transition kernel $Q$ and set $k=k+1$. Repeat this procedure until a jump time $j_{k+1}\geq T$ is generated and set $j_{k+1}=T$. These steps are summarised in Algorithm \ref{alg:PDifMP_constantLambda} in Appendix \ref{appA.1}.

\begin{figure}[t]
    \centering
    \includegraphics[width=\textwidth]{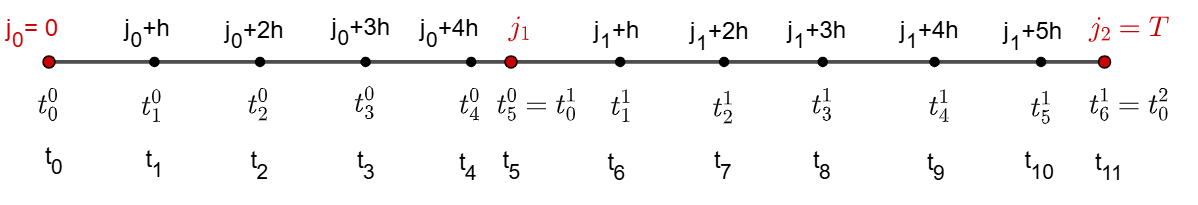}
    \caption{Example time grid for $N^j=3$ and $N^x=12$.}
    \label{fig:adaptiveGrid_totalGrid}
\end{figure}

\paragraph{Simulation of PDifMP paths with process-dependent jump rate function}

In this case, the diffusion dynamics remain unchanged, while the jump times can no longer be drawn directly from an exponential distribution and should instead be simulated using the \textit{thinning method}~\cite{buckwar2025numerical,lewis1979simulation}, which adapts the classical approach for non-homogeneous Poisson processes to state-dependent intensities. We consider non-constant jump rate functions $\Lambda$, which may depend on the states of the process $U_t=(X_t, Z_t)$ and are bounded by a constant $\lambda > 0$. In what follows, we restrict attention to the case where $\Lambda$ depends only on the continuous component, i.e, $\Lambda(U_t):=\Lambda(X_t)$.

The steps are as follow: Draw a candidate $j_{new}$ for the next jump time from an exponential distribution with rate $\lambda$ and simulate the path $x$ of $X_t$ up to $j_{new}$ (cf. Algorithm \ref{alg:PathFlowPhi}). The candidate $j_{new}$ is accepted as a jump time with probability $\Lambda(x_{j_{new}})/\lambda$. If accepted, the new discrete state $z_{k+1}$ is drawn from the transition kernel $Q$ and the index $k$ is incremented by one. Regardless of whether the candidate was accepted or not, we set $j_{old} = j_{new}$, sample a new waiting time $\tau$ from the exponential distribution with rate $\lambda$ and add it to the previous candidate $j_{old}$ in order to obtain the new candidate $j_{new}$. The whole simulation procedure is reported in Algorithm \ref{alg:PDifMP_boundedLambda} in Appendix \ref{appA.1}.

Note that the time grid for the path $x$ is not specified explicitly in Algorithms \ref{alg:PDifMP_constantLambda} and \ref{alg:PDifMP_boundedLambda}. We denote the points on this grid by $t_i$, $i=0,\ldots,N^x-1$, where $N^x$ is the total number of time points. An example configuration with $N^x=12$ and $N^j=3$ jumps is shown in Figure \ref{fig:adaptiveGrid_totalGrid}, which also illustrates how this global grid relates to the piecewise grids reported in Figure \ref{fig:adaptiveGrid}.


\section{Representative example PDifMPs}
\label{sec:3:PDifMPs}

In this section we introduce diverse example PDifMPs that will serve as test problems for the inference via ABC. Each PDifMP test problem is motivated by an underlying SDE, which governs the dynamics of the component $X$ between two consecutive jumps. A suitable transition kernel is then specified for each test problem to determine the values $z_k$ of the component $Z$, which strongly influences the dynamics of $X$. In most analyses in Section \ref{sec:5:Inference}, we focus on a constant jump rate function $\Lambda\equiv \lambda > 0$. In Section \ref{sec:5:2:NonConstantLambda}, however, we extend our analysis to non-constant jump rate functions that depend on the state of the component $X$. 

For the reader's convenience, we first recall the motivating SDEs in Section \ref{sec:3:1:SDEs} and then use them to construct the corresponding PDifMPs in Section \ref{sec:3:2:testProblems}.

\subsection{Underlying SDEs}
\label{sec:3:1:SDEs}

\paragraph{Wiener process with drift (WPWD)}

This process is described by the SDE
\begin{equation}
\label{eq:SDE_WPWD}
    dX_t = \gamma dt + \sigma dW_t, \quad t \in [0,T], \quad X_0 = x_0 \in \mathbb{R},
\end{equation}
with drift $\gamma \in \mathbb{R}$ and noise parameter $\sigma > 0$. Its solution (flow) is given by
\begin{equation*}
    X_t = \phi(t,x_0;\gamma) = x_0 + \gamma t + \sigma W_t.
\end{equation*}
Note that, for each $t \in [0,T]$,
\begin{equation*}
    X_t \sim \mathcal{N}\big(x_0+\gamma t,\,\sigma^2 t\big).
\end{equation*}
Therefore, since both the mean and variance of this process change linearly with time, its distribution does not converge to a stationary distribution.

\paragraph{Ornstein-Uhlenbeck process (OU)}

This process is described by the SDE
\begin{equation}
\label{eq:SDE_OU}
    dX_t = \gamma_1 (\gamma_2 - X_t)dt + \sigma d W_t, \quad t\in[0,T], \quad X_0 = x_0 \in \mathbb{R},
\end{equation}
where $\gamma_2 \in \mathbb{R}$ and $\gamma_1,\sigma > 0$. 
This SDE can be solved explicitly. Its solution (flow) is given by
\begin{equation*}
    X_t = \phi(t,x_0;\gamma_1,\gamma_2) = x_0e^{-\gamma_1 t} + \gamma_2(1-e^{- \gamma_1 t}) + \sigma\int_0^t e^{-\gamma_1(t-s)}dW_s.
\end{equation*}
For each $t \in [0,T]$,
\begin{equation*}
    X_t \sim \mathcal{N}\Bigl(x_0e^{-\gamma_1 t} + \gamma_2\left(1-e^{-\gamma_1 t}\right),\,\frac{\sigma^2}{2\gamma_1}\left(1-e^{-2 \gamma_1 t}\right)\Bigr),
\end{equation*}
where the variance can be determined using It\^o's isometry. Therefore, for $t\to \infty$, the distribution of the OU process converges to a stationary distribution given by
\begin{equation}
\label{eq:stat_mean_OU}
    \mathcal{N}\left(\gamma_2,\, \frac{\sigma^2}{2\gamma_1}\right).
\end{equation}

\paragraph{Weakly damped stochastic harmonic oscillator (WDSHO)}

This process is two-dimensional, i.e. $X_t=(X_t^{(1)},X_t^{(2)})^\top$, $t\in [0,T]$, and is described by the SDE
\begin{equation}
\label{eq:SDE_WDSHO}
    d\binom{X_t^{(1)}}{X_t^{(2)}} = \binom{X_t^{(2)}}{-\gamma_1^2 X_t^{(1)} - 2\gamma_2 X_t^{(2)}}dt + \binom{0}{\sigma} dW_t, \quad t \in [0,T], \quad X_0 = x_0 = (x_0^{(1)},x_0^{(2)})^\top,
\end{equation}
where $\gamma_1 ,\gamma_2 > 0$ with $\gamma_1^2 - \gamma_2^2 > 0$ and $\sigma>0$. The equation is a linear SDE with additive noise and can be solved explicitly. Specifically, it can be written as
\begin{equation*}
    dX_t = A(\gamma_1,\gamma_2) X_tdt + \Sigma dW_t, \quad t \in [0,T], \quad X_0 = x_0 = (x_0^{(1)},x_0^{(2)})^\top,
\end{equation*}
where
\begin{equation}
\label{eq:Az_Sigma}
    A(\gamma_1,\gamma_2):=\begin{pmatrix}
    0 & 1 \\
    -\gamma_1^2 & -2\gamma_2
    \end{pmatrix} \quad \text{and} \quad \Sigma := \binom{0}{\sigma},
\end{equation}
with solution (flow) given by
\begin{equation*}
    X_t=\phi(t,x_0;\gamma_1,\gamma_2)=e^{A(\gamma_1,\gamma_2)t}x_0 + \int_0^t e^{A(\gamma_1,\gamma_2)(t-s)}\Sigma \ dW_s.
\end{equation*}
Therefore, for each $t \in [0,T]$, 
\begin{equation*}
    X_t \sim \mathcal{N}\Big(\textrm{Exp}(t;\gamma_1,\gamma_2)x_0 ,\, \textrm{Cov}(t;\gamma_1,\gamma_2) \Big),
\end{equation*}
where 
\begin{equation*}
    \textrm{Exp}(t;\gamma_1,\gamma_2)=e^{A(\gamma_1,\gamma_2)t}= e^{- \gamma_2 t} \begin{pmatrix}
        \cos(\kappa t) + \frac{\gamma_2}{\kappa} \sin (\kappa t) & \frac{1}{\kappa} \sin (\kappa t) \\
        -\frac{\gamma_1^2}{\kappa} \sin (\kappa t) & \cos (\kappa t)-\frac{\gamma_2}{\kappa} \sin (\kappa t)
    \end{pmatrix},
\end{equation*}
with $\kappa:=\sqrt{\gamma_1^2 - \gamma_2^2}>0$, and
\begin{equation}
\label{eq:cov_weaklyDamped}
    \textrm{Cov}(t;\gamma_1,\gamma_2) = \int_0^t e^{A(\gamma_1,\gamma_2)(t-s)}\Sigma\Sigma^{\top}(e^{A(\gamma_1,\gamma_2)(t-s)})^{\top} ds,
\end{equation}
with
\begin{align*}
    \textrm{Cov}(t;\gamma_1,\gamma_2)_{11} & =  \frac{\sigma^2}{4\gamma_2 \gamma_1^2} - \frac{\sigma^2 e^{-2\gamma_2t}}{4\gamma_2 \gamma_1^2 \kappa^2} \left(\gamma_1^2 -\gamma_2^2\cos(2\kappa t) + \gamma_2 \kappa \sin(2\kappa t)\right),\\    
    \textrm{Cov}(t;\gamma_1,\gamma_2)_{12} = \textrm{Cov}(t;\gamma_1;\gamma_2)_{21} & = \frac{\sigma^2}{2 \kappa^2} e^{-2\gamma_2t}\sin^2(\kappa t), \\
    \textrm{Cov}(t;\gamma_1,\gamma_2)_{22} & = \frac{\sigma^2}{4\gamma_2} - \frac{\sigma^2 e^{-2\gamma_2t}}{4\gamma_2\kappa^2}\left( \gamma_1^2 - \gamma_2^2\cos(2\kappa t) - \gamma_2 \kappa \sin(2\kappa t)\right). 
\end{align*}

It follows that, for $t \to \infty$, the distribution of this process converges to a stationary distribution given by
\begin{equation*}
    \mathcal{N}\Bigl(\binom{0}{0},\begin{pmatrix}
    \frac{\sigma^2}{4\gamma_2\gamma_1^2} & 0 \\
    0 & \frac{\sigma^2}{4\gamma_2}
\end{pmatrix} \Bigr).
\end{equation*}
Thus, the value of the parameter $\gamma_2$ influences the variance of the stationary distribution of both components of the process. In contrast, the value of $\gamma_1$ only influences the variance of the stationary distribution of the first component. Moreover, $\gamma_1$ impacts the oscillation frequencies of both components. In particular, larger values of $\gamma_1$ correspond to smaller variance and higher frequency in the first component, and to higher frequency in the second component.

\paragraph{Simple stochastic harmonic oscillator (SimpleSHO)}

This process corresponds to a stochastic harmonic oscillator with $\gamma_2\equiv 0$. Specifically, it is described by SDE \eqref{eq:SDE_WDSHO} with $\gamma_2$ set to zero, which can be written as
\begin{equation*}
    dX_t = A(\gamma_1,0) X_tdt + \Sigma dW_t, \quad t \in [0,T], \quad X_0 = x_0 = (x_0^{(1)},x_0^{(2)})^\top,
\end{equation*}
where $A$ and $\Sigma$ are as in \eqref{eq:Az_Sigma}. 
Its solution (flow) is given by
\begin{equation*}
    X_t=\phi(t,x_0;\gamma_1)=e^{A(\gamma_1,0)t}x_0 + \int_0^t e^{A(\gamma_1,0)(t-s)}\Sigma \ dW_s.
\end{equation*}
Thus, for each $t \in [0,T]$, 
\begin{equation*}
    X_t \sim \mathcal{N}\Big(\textrm{Exp}(t;\gamma_1)x_0 , \,\textrm{Cov}(t;\gamma_1) \Big),
\end{equation*}
where 
\begin{equation*}
    \textrm{Exp}(t;\gamma_1)=e^{A(\gamma_1,0)t}= \begin{pmatrix}
        \cos(\gamma_1t) & \frac{1}{\gamma_1}\sin(\gamma_1t) \\
        -\gamma_1\sin(\gamma_1 t) & \cos(\gamma_1t)
    \end{pmatrix},
\end{equation*}
and
\begin{equation}
\label{eq:cov_simple}
    \textrm{Cov}(t;\gamma_1)= \int_0^t e^{A(\gamma_1,0)(t-s)}\Sigma\Sigma^{\top}(e^{A(\gamma_1,0)(t-s)})^{\top} ds = \frac{\sigma^2}{2}
    \begin{pmatrix}
        \frac{2\gamma_1t - \sin(2\gamma_1 t)}{2\gamma_1^3} & \frac{\sin^2(\gamma_1 t)}{\gamma_1^2}  \\
        \frac{\sin^2(\gamma_1 t)}{\gamma_1^2} & \frac{2\gamma_1t + \sin(2\gamma_1t)}{2\gamma_1}
    \end{pmatrix}.
\end{equation}
Thus, unlike the WDSHO, the SimpleSHO does not converge to a stationary distribution.

\subsection{PDifMP test problems}
\label{sec:3:2:testProblems}

We now introduce four representative PDifMP test problems (TPs), each constructed by extending one of the SDEs from Section~\ref{sec:3:1:SDEs} into a hybrid stochastic framework. In all cases, a deterministic model parameter of the underlying SDE, typically $\gamma$, $\gamma_1$, or $\gamma_2$, is replaced by a piecewise-constant jump process $Z$, which evolves at random jump times. This construction couples the underlying continuous dynamics of $X$ with the discrete switching behaviour of $Z$. 

\begin{remark}
    Recall that, in general, $Q : E \times \mathcal{B}(D_2) \to [0,1]$ denotes a transition kernel assigning to each $u=(x,z)\in E$ a probability measure $Q(u,\cdot)$ on $(D_2,\mathcal{B}(D_2))$. In all test problems below, $Q$ is deterministic and thus reduces to a measurable mapping $Q : E \to D_2$, assigning a single post-jump value $z_{k+1}$ rather than a distribution. We therefore write $z_{k+1} = Q(x_{J_{k+1}})$ or $z_{k+1} = Q(z_k)$, depending on the case considered.
\end{remark}

\paragraph{TP~1 (OU-PDifMP)} The idea of the example PDifMP, which is based on the Ornstein-Uhlenbeck process \eqref{eq:SDE_OU}, is to let the mean $\gamma_2$ of the stationary distribution (cf. \eqref{eq:stat_mean_OU}) change at random jump times, i.e. $\gamma_2$ is replaced by a jump process $Z$. Specifically, let $b > 0$, then a value of $z_k$ is either equal to $b$ (positive) or equal to $-b$ (negative), depending on the position $x_{J_k}$ of the process $X$ at the jump time $J_k$. Therefore, the state space of the PDifMP is $E = D_1 \times D_2 = \mathbb{R} \times \{-b,b\}$. In the following, we set $x_0=0$ and $z_0=b$. Moreover, to simplify the notation (in particular, to remove the now unnecessary subscript), we replace $\gamma_1$ with $\eta$. 

Structurally, the OU-PDifMP comprises the following components:   

\begin{itemize}
    \item Underlying SDE: 
    \begin{align*}
        dX_t &= \eta (z_k - X_t)dt + \sigma d W_t, \quad t \in [J_k,J_{k+1}), \\
        X_{J_k} &= x_{J_k} \in D_1 = \mathbb{R},
    \end{align*}
     with  stochastic flow map given by
     \begin{equation*}
         X_t = \phi(t-J_k,x_{J_k},z_k) = x_{J_k}e^{-\eta(t-J_k)} + z_k(1-e^{-\eta(t-J_k)}) + \sigma \int_{J_k}^t e^{-\eta(t-s)}dW_s.
     \end{equation*}
    
    \item Transition function:  $$z_{k+1} = Q(x_{J_{k+1}})= \begin{cases}
        b, & x_{J_{k+1}} \leq 0 \\
        -b, & x_{J_{k+1}} > 0 \\
    \end{cases},$$
    where $x_{J_{k+1}}=\phi(J_{k+1}-J_{k},x_{J_k},z_k)$.
\end{itemize}

\paragraph{TP~2 (WDSHO-PDifMP)} The idea of the example PDifMP, which is based on the weakly damped stochastic harmonic oscillator \eqref{eq:SDE_WDSHO}, is to let the frequency of the described oscillations change at the random jump times. Specifically, $\gamma_2$ is now denoted by $\eta$ and $\gamma_1$ is replaced by a jump process $Z$ with values $z_k$ that alternate between a fixed number $2$ and a parameter $b>\eta$. The state space of this PDifMP is given by $E = D_1 \times D_2 = \mathbb{R}^2 \times \{2,b\}$, the transition mapping for the $z_k$-values in $D_2$ is independent from~$X$, and the initial values for the diffusion and jump component are set to  $x_0 = (1,1)^\top$ and $z_0 = b$, respectively.

In particular, the WDSHO-PDifMP comprises the following parts:
\begin{itemize}
    \item Underlying SDE: 
    \begin{align*}
        d\binom{X_t^{(1)}}{X_t^{(2)}} &= \binom{X_t^{(2)}}{-z_k^2 X_t^{(1)} - 2 \eta X_t^{(2)}}dt + \binom{0}{\sigma} dW_t, \quad t \in [J_k,J_{k+1}), \\
        X_{J_k} &= x_{J_k} \in D_1 = \mathbb{R}^2,
    \end{align*}
    with solution
    \[X_t = \phi(t-J_k,x_{J_k},z_k) = e^{A(z_k,\eta)(t-J_k)} x_{J_k} + \int_{J_k}^te^{A(z_k,\eta)(t-s)}\Sigma \ dW_s,\]
    where $A$ and $\Sigma$ are as in \eqref{eq:Az_Sigma}.
    
    \item Transition function: $$z_{k+1} = Q(z_k)= \begin{cases}
        2, & z_{k} = b \\
        b, & \text{else}
    \end{cases}.$$
\end{itemize}

\paragraph{TP~3 (WPWD-PDifMP)} The idea of the example PDifMP, which is based on the Wiener process with drift \eqref{eq:SDE_WPWD}, is to let the  drift $\gamma$ change at random jump times, by replacing it with a jump process $Z$. Specifically, let $b > 0$, then a drift value $z_k$ is either equal to $b$ (positive) or equal to $-b$ (negative), depending on the position $x_{J_k}$ of the process $X$ at the jump time $J_k$ (as for the OU-PDifMP). The state space of this PDifMP is $E = D_1 \times D_2 = \mathbb{R} \times \{-b,b\}$, and the initial values are set to $x_0 = 0$ and $z_0 = b$.

The WPWD-PDifMP can be described by:
\begin{itemize}
    \item Underlying SDE: 
    \begin{align*}
        dX_t &= z_k dt + \sigma dW_t, \quad t \in [J_k,J_{k+1}), \\
        X_{J_k} &= x_{J_k} \in D_1 = \mathbb{R},
    \end{align*}
    with solution
    \begin{equation*}
        X_t = \phi(t-J_k,x_{J_k},z_k) = x_{J_k} + z_k(t-J_k) + \sigma (W_t - W_{J_k}).
    \end{equation*}
    \item Transition function: $$z_{k+1} = Q(x_{J_{k+1}}) = \begin{cases}
        b, & x_{J_{k+1}} \leq 0 \\
        -b, & x_{J_{k+1}} > 0 \\
    \end{cases},$$
    where $x_{J_{k+1}}=\phi(J_{k+1}-J_{k},x_{J_k},z_k)$.
\end{itemize}

\paragraph{TP~4 (SwitchedSHO-PDifMP)} This example PDifMP is based on a combination of both the weakly damped and the simple stochastic harmonic oscillator (cf. Eq. \eqref{eq:SDE_WDSHO}). Specifically, $\gamma_1$ is now denoted by $\eta$ and $\gamma_2$ is replaced by a jump process $Z$ with values $z_k$ that alternate between $0$ (SimpleSHO case) and a parameter $b \in (0,\eta)$ (WDSHO case). Similar as in TP~2, the transition function is independent of $X$, the state space is $E=D_1\times D_2=\mathbb{R}^2 \times \{0,b\}$, and the initial values are set to $x_0=(1,1)^\top$ and $z_0=b$.

The PDifMP can be described by:
\begin{itemize}
    \item Underlying SDE: 
    \begin{align*}
        d\binom{X_t^{(1)}}{X_t^{(2)}} &= \binom{X_t^{(2)}}{-\eta^2 X_t^{(1)} - 2z_k X_t^{(2)}}dt + \binom{0}{\sigma} dW_t, \quad t \in [J_k,J_{k+1}), \\
        X_{J_k} &= x_{J_k} \in D_1 = \mathbb{R}^2,
    \end{align*}
    with solution
    \[X_t = \phi(t-J_k,x_{J_k},z_k) = e^{A(\eta,z_k)(t-J_k)} x_{J_k} + \int_{J_k}^te^{A(\eta,z_k)(t-s)}\Sigma \ dW_s,\]
    where $A$ and $\Sigma$ are as in \eqref{eq:Az_Sigma}.
    
    \item Transition function: $$z_{k+1} = Q(z_k) = \begin{cases}
        0, & z_k = b \\
        b, & \text{else}
    \end{cases}.$$
\end{itemize}

\begin{remark}
    In TP~$1$ and TP~$3$ we allow so-called \textit{no-move jumps}, i.e. the transition function is defined such that the next value $z_{k+1}$ may equal the previous value $z_k$. These no-move jumps are not visible in the path of the PDifMP, since no regime change occurs at the corresponding jump times. 
\end{remark}

\begin{remark}\label{rem:extensions}
    Several extensions of the PDifMP framework considered here are possible, although they are beyond the scope of this manuscript. First, note that in all TPs discussed, the underlying SDEs are explicitly solvable, which allows for exact simulation of the corresponding flow $\phi$ (cf. Algorithm \ref{alg:PathFlowPhi} in Appendix \ref{appA.1}). In many practical applications, however, the drift coefficient may be nonlinear, making the SDE unsolvable in closed form. In such cases, appropriate numerical methods can be employed to simulate approximate realisations of the flow $\phi$, cf.~\cite{buckwar2025numerical}. 
    
    Second, in all examples considered, the state space $D_2$ of the component $Z$ is binary. This setting can be extended to a continuous state space, for instance $D_2=\mathbb{R}$.  In this case, the next value $z_{k+1}$ can be determined through a suitable stochastic transition mechanism, for instance by sampling from a continuous distribution that may depend on the current state of the process, e.g., a Gaussian transition kernel centred at $z_k$, with a variance that may depend on the current position of the process. 
\end{remark}

\begin{remark}
    Note that, the adaptive time grid of PDifMPs (cf. Figures \ref{fig:adaptiveGrid} and \ref{fig:adaptiveGrid_totalGrid}) may result in very small step sizes, which can introduce numerical inaccuracies in the computation of the covariance matrices  \eqref{eq:cov_weaklyDamped} and \eqref{eq:cov_simple}. To prevent such issues, we round the jump times in TP 2 and TP 4 to guarantee a minimal step size. 
\end{remark}

\paragraph{Path of test problems}

\begin{figure}[t]
    \centering
    \includegraphics[width=0.88\textwidth]{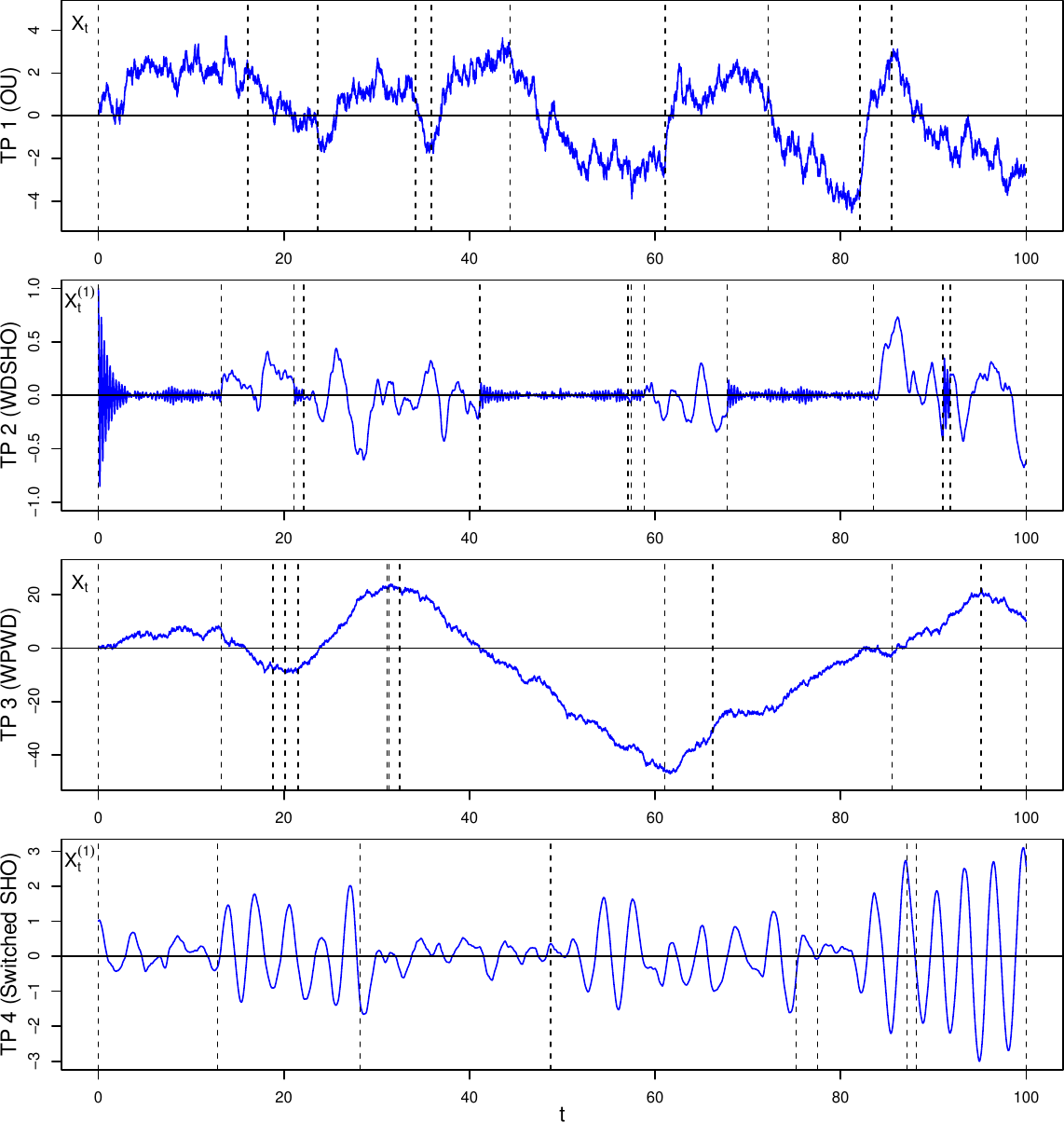}
    \caption{Paths of the four test problems.}
    \label{fig:paths}
\end{figure}

Figure \ref{fig:paths} shows a sample path of $X$ over the time interval $[0,100]$ for each of the four considered PDifMP test problems. The vertical dashed lines indicate the corresponding jump times. For TP~2 (WDSHO-PDifMP) and TP~4 (SwitchedSHO-PDifMP) only the path of the first component $X^{(1)}$ is reported.

The path of TP~1 (OU-PDifMP) illustrates how the component $X$ alternates between the two regimes, with ``stationary'' means $z_k=b=2$ and $z_k=-b=-2$, respectively. 

The path of TP~2 (WDSHO-PDifMP) also switches between two regimes. In the first regime, it exhibits a high frequency and small variance ($z_k=b=20$), while in the second regime, it shows a lower frequency and larger variance ($z_k=2$). In both regimes, the process fluctuates around the underlying stationary mean $0$. 

The path of TP~3 (WPWD-PDifMP) illustrates how the component $X$ evolves as the drift alternates between $z_k=b=2$ and $z_k=-b=-2$.

The path of TP~4 (SwitchedSHO-PDifMP) demonstrates how the component $X$ alternates between two types of stochastic harmonic oscillators: a weakly damped oscillator ($z_k=b=1$) with underlying stationary distribution $\mathcal{N}(0,1/16)$ and a non-ergodic simple oscillator ($z_k=0$) whose amplitude, on average, increases over time.

\paragraph{Ergodic behaviour of test problems}

All considered example PDifMPs exhibit empirically observed ergodic behaviour. For the $X$ component of these processes, this means that time averages computed along a single sufficiently long trajectory coincide with ensemble (phase-space) averages taken over multiple realisations. Formally, 
\begin{equation*}
    \lim_{T \to +\infty} \frac{1}{T} \int_0^T f(X_s)ds = \int f(x) \mu(dx),
\end{equation*}
i.e., the long-time average of a measurable function $f$ converges to its expectation with respect to the underlying invariant measure $\mu$. 

This behaviour is crucial for parameter inference in PDifMPs via ABC (cf. Sections~\ref{sec:4:ABC} and \ref{sec:5:Inference}), as it allows reliable statistical information to be extracted from a single sufficiently long trajectory. In particular, trajectories generated under the same parameter values should yield similar summary statistics, whereas different parameter values should produce sufficiently distinct summaries. In the absence of ergodic behaviour, the within-parameter variability of the summaries may become comparable to, or larger than, the differences induced by changing the parameters, preventing ABC from reliably distinguishing between parameter values.

\begin{figure}[t]
    \centering
    \includegraphics[width=\textwidth]{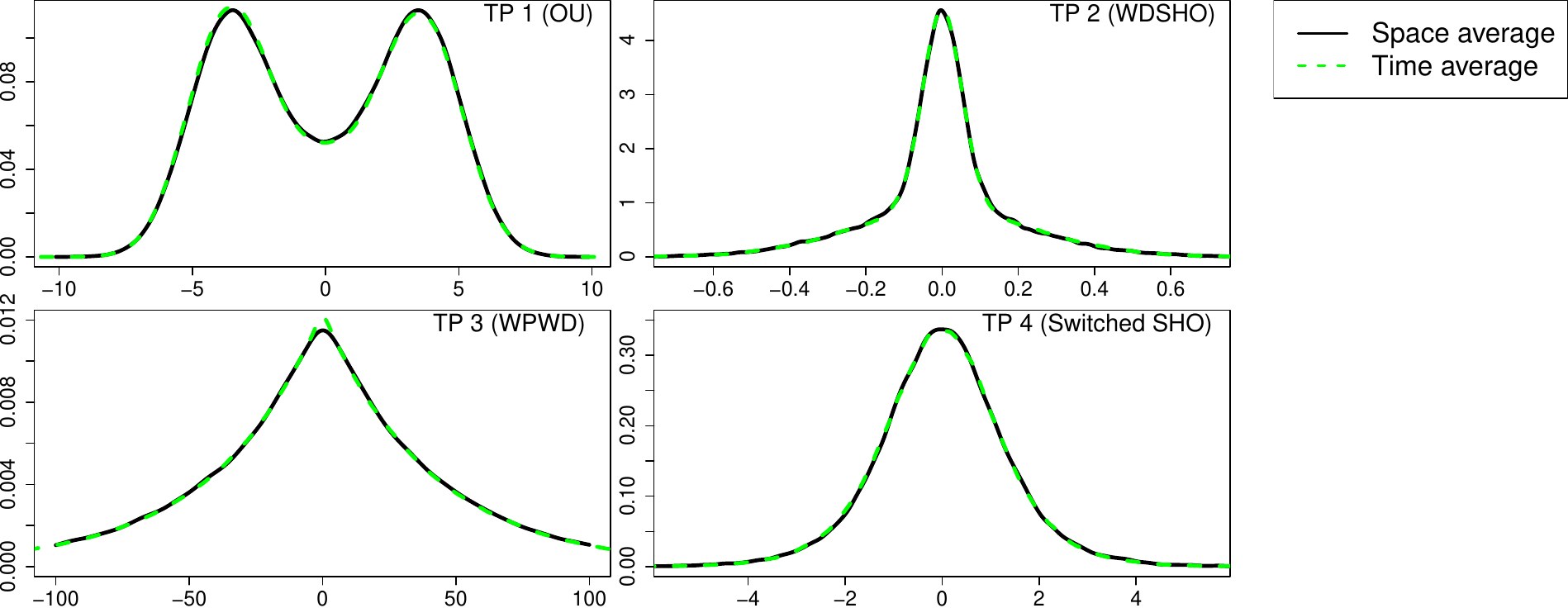}
    \caption{Space and time averages of the four test problems.}
    \label{fig:space_time}
\end{figure}

Figure \ref{fig:space_time} illustrates the empirically observed ergodic behaviour for the considered PDifMP test problems, showing that densities derived from a single path (green dashed lines) closely match densities computed from multiple realisations of $X_{t^*}$, $t^* \gg 0$ (black solid lines). Note that the SDEs underlying the first two test problems are ergodic, so the observed ergodic behaviour of the resulting PDifMPs might be expected. In contrast, TP3 is based entirely on non-ergodic SDE dynamics, while TP4 switches between ergodic and non-ergodic SDE dynamics, making the emergence of ergodic behaviour in these cases particularly remarkable. 

The density of TP~1 (OU-PDifMP) is bimodal, exhibiting two different means. Similarly, TP~2 (WDSHO-PDifMP) shows two different variances: a tall, narrow central peak (resulting from the smaller variance) and wide, heavy tails (resulting from the larger variance). Moreover, the peaks of the densities for TP~3  (WPWD-PDifMP) are not perfectly aligned, raising some doubts regarding the theoretical ergodicity of that test problem.

A key factor distinguishing TP3 (WPWD-PDifMP) from the other test problems is that its underlying continuous SDE completely lacks ergodicity. Consequently, the empirically observed ergodic behaviour of this PDifMP relies entirely on the stabilising effect of the jump dynamics. When jumps and effective regime changes occur infrequently, the process can spend long periods evolving according to one of the underlying SDE regimes, which do not possess stationary distributions. This behaviour is also reflected in the sample paths (Figure~\ref{fig:paths}), where TP3 spans a considerably wider spatial range than the other test problems.

\begin{figure}[t]
    \centering
    \includegraphics[width=0.4\textwidth]{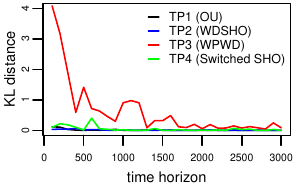}
    \caption{KL distance between the density estimates based on the space and time averages as a function of the time horizon $T$ for the four test problems.}
    \label{fig:KL_distance}
\end{figure}

To investigate the agreement between space and time averages more closely, Figure~\ref{fig:KL_distance} shows the Kullback--Leibler (KL) distance between the corresponding densities as a function of the time horizon used to compute the time average for all four test problems. For this comparison, the time horizon and the number of realisations used to compute the reference space average were kept fixed, while only the time horizon used to compute the time average was varied. For TP3, a considerably longer time horizon is required before the KL distance becomes small. Moreover, within the considered range of time horizons, the KL distance fluctuates around a small positive value rather than showing a clear tendency towards zero. This may be due to the finite time horizon and finite number of realisations used to approximate the reference space average, or it may reflect the possibility that TP3 is not theoretically ergodic, as mentioned above.

We further examine whether the agreement between space and time averages observed for the parameter configurations in Figure~\ref{fig:space_time} persists across a broader parameter region of interest. For TP1 (OU-PDifMP), TP2 (WDSHO-PDifMP), and TP4 (SwitchedSHO-PDifMP), the resulting KL distances remain consistently small across the considered parameter configurations. For TP3 (WPWD-PDifMP), the detailed results across parameter configurations are presented in Table~\ref{fig:tp3_full_analysis}. Larger KL distances occur mainly for very small values of the jump rate $\lambda$. As indicated by the time-horizon analysis above, these larger discrepancies may be due to the considered time horizons being insufficient to obtain small KL distances. Moreover, very small jump rates correspond to settings in which regime transitions occur only rarely and are therefore of limited relevance for the PDifMP dynamics considered here. For moderate to high jump rates, the discrepancies between space and time averages remain small.

\begin{table}[htbp]
  \centering
  \begin{minipage}{0.48\textwidth}
    \centering
    \small
    \begin{tabular}{|c|ccccc|}
      \hline
      $\lambda$\textbackslash $\sigma$& 0.1 & 2.5 & 5 & 7.5 & 10 \\
      \hline
      0.001 & \textbf{7.365} & \textbf{3.531} & \textbf{3.113} & \textbf{3.901} & \textbf{3.063} \\
      0.25  & 0.052 & 0.026 & 0.028 & 0.021 & 0.029 \\
      0.5   & 0.004 & 0.009 & 0.017 & 0.015 & 0.016 \\
      0.75  & 0.021 & 0.006 & 0.008 & 0.007 & 0.017 \\
      1     & 0.003 & 0.007 & 0.009 & 0.012 & 0.018 \\
      \hline
    \end{tabular}
    \\[2mm]
    \small (a) KL distances for $\lambda$ vs. $\sigma$
    \label{sub:tp3_tab_sigma}
  \end{minipage}%
  \hfill
  \begin{minipage}{0.48\textwidth}
    \centering
    \small
    \begin{tabular}{|c|ccccc|}
      \hline
      $\lambda$\textbackslash $b$& 0.1 & 2.5 & 5 & 7.5 & 10 \\
      \hline
      0.001 & \textbf{6.347} & \textbf{4.683} & \textbf{5.454} & \textbf{5.379} & \textbf{6.225} \\
      0.25  & 0.667 & 0.023 & 0.006 & 0.026 & 0.014 \\
      0.5   & 0.173 & 0.009 & 0.007 & 0.027 & 0.005 \\
      0.75  & 0.024 & 0.010 & 0.004 & 0.019 & 0.021 \\
      1     & 0.062 & 0.010 & 0.006 & 0.002 & 0.006 \\
      \hline
    \end{tabular}
    \\[2mm]
    \small (b) KL distances for $\lambda$ vs. $b$
    \label{sub:tp3_tab_b}
  \end{minipage}

  \vspace{3mm}
  \caption{Sensitivity of KL distance between space and time average to parameters $\lambda$, $\sigma$, and $b$ for TP3 (WPWD-PDifMP).}
  \label{fig:tp3_full_analysis}
\end{table}


\section{Approximate Bayesian computation}
\label{sec:4:ABC}

We now address the problem of parameter inference within PDifMPs. For complex stochastic models such as PDifMPs, the underlying likelihood function is difficult to handle. Even if the underlying SDE of a PDifMP is exactly solvable and its transition density known, constructing a likelihood for the full PDifMP remains unclear. In principle, one could attempt a piecewise formulation if the exact jump times were known, but this is rarely the case in practice. When the underlying SDE is not analytically solvable or the PDifMP is only partially observed, even such piecewise constructions become infeasible. Approximate Bayesian computation (ABC) addresses this challenge by replacing the unavailable likelihood with extensive simulations of synthetic datasets generated from the model. These simulated datasets are then used to construct an approximate posterior distribution for the unknown parameter vector of interest.

Specifically, let $y$ denote the observed dataset, $\theta$ the parameter vector to be inferred from $y$, and $\pi(\theta)$ a prior distribution for $\theta$. The basic ABC algorithm proceeds as follows:
\begin{enumerate}
    \item Draw a parameter value $\theta^{'}$ from the prior distribution $\pi(\theta)$.
    \item Conditional on $\theta^{'}$, simulate a synthetic dataset $y_{\theta^{'}}$ from the model.
    \item Accept $\theta^{'}$ as a sample from the ABC posterior distribution if the distance $d(\cdot,\cdot)$ between a set of summary statistics $s(\cdot)$ of the observed data $y$ and the synthetic dataset $y_{\theta^{'}}$ is below a chosen threshold $\delta \geq 0$, i.e. if $d(s(y),s(y_{\theta^{'}}))<\delta$.
\end{enumerate}
These steps are repeated until a desired number of accepted samples from the ABC posterior distribution has been obtained. 

\subsection{ABC for PDifMPs}

In this section, the ABC methodology is adapted for parameter inference in PDifMPs.

\paragraph{Observed dataset}

For PDifMPs, the observed dataset $y$, on which inference will be performed, is defined as
\begin{equation}
\label{eq:observed_y}
    y:=\{x,N^j\}, 
\end{equation}
where $x=(x_{t_i})_{i=0}^{N^x-1}$ is a time-series dataset consisting of discrete-time measurements of the component $X$ at the times $t_i$, $i=0,\ldots,N^x-1$, with $t_0=0$ and $t_{N^x-1}=T$, and $N^j$ is the number of jumps that occurred within the observation interval $[0,T]$.

When $X$ has multiple components, as in TP~2 (WDSHO-PDifMP) and TP~4 (SwitchedSHO-PDifMP), we assume that only the first coordinate $X^{(1)}$ is observed. That is, the observed time series is of the form $x=x^{(1)}=(x^{(1)}_{t_i})_{i=0}^{N^x-1}$. This allows the proposed ABC method to handle challenging partially observed cases, as may be relevant in real-world settings. We note, however, that observing the second component $X^{(2)}$ or both components of $X$ would be possible and would lead to comparable inferential results.

Importantly, assuming that $N^j$ is observed does not imply that the jump process $Z$ or the individual jump times are fully observed. Rather, we consider an observation setting in which jump events are recorded separately from the continuous state trajectory. Such settings arise, for example, in reliability and repairable-system applications, where event counts and failure times are commonly available as observed data for statistical inference \cite{guo2010piecewise}.
Our general inference framework therefore does not rely on observing the jump times $j=(j_k)_{k=0}^{N^j-1}$, which are often not available in practice. In some applications, however, the jump times may indeed be observed. To illustrate a case where the exact locations of jump times provide additional information for inference, an example is presented in Section~\ref{sec:6:TP3}. Moreover, a robustness analysis of the inferential results under misspecification of the observed jump count $N^j$ is provided in Appendix \ref{app:A:Robustness}.

Furthermore, it would be possible to include the second jump-related component $z = (z_k)_{k=1}^{N^j}$ in the inference. However, since $z$ might not be observable in practice, we do not use it in our analysis. 

\paragraph{Parameter vector}

The example PDifMPs introduced in the previous section are defined by an underlying SDE, a  transition function, and a rate function governing the jump dynamics. Our goal is to simultaneously infer one parameter for each component of the model. Specifically, the SDE is influenced by a diffusion parameter $\sigma$, the transition function depends on a parameter $b$ (which in turn affects the SDE dynamics), and the jump rate function is determined by a parameter $\lambda$. We aim to infer these three parameters jointly, i.e., to estimate 
\begin{equation}
\label{eq:theta}
	\theta = (\sigma,b,\lambda).
\end{equation}

Note that, in most of our analyses, we fix $\eta$ (additional drift parameter of the underlying SDE) to $1/2$, $1$, and $2$ for TP~1 (OU-PDifMP), TP~2 (WDSHO-PDifMP), and TP~4 (SwitchedSHO-PDifMP), respectively. However, in Section \ref{sec:5.3_4parameter}, we include $\eta$ in the parameter vector $\theta$ and demonstrate that this additional parameter can be also inferred without difficulty.

\paragraph{Prior distribution}

Following \cite{Buckwar2020,Ditlevsenetal2023,SAMSON2025108095}, we employ uniform prior distributions. In particular, we consider
\begin{equation}
\label{eq:priors}
    \sigma \sim U(0,10), \quad b \sim U(0,10), \quad \lambda \sim U(0,1),
\end{equation}
with exceptions for specific test problems. For TP~2 (WDSHO-PDifMP), we choose $b~\sim~U(2,100)$, since the model is only well-defined for $b > \eta$, which is set to $1$. For TP~4 (SwitchedSHO-PDifMP), we consider $b \sim U(0,1)$, as the problem is only well-defined for $b \in (0,\eta)$, and $\eta$ is set to $2$. Note that our experiments show that the choice of the prior has limited impact on the inferential results, consistent with the findings of \cite{SAMSON2025108095} for ABC inference in SDE models.

\paragraph{Model simulation} 

A key step in the ABC methodology is the simulation of synthetic datasets~$y_{\theta}$ from the model. For PDifMPs, we employ the two algorithms introduced in Section \ref{sec:2:PDifMPs} (see also Appendix \ref{appA.1}): Algorithm~\ref{alg:PDifMP_constantLambda} for constant jump rates, and Algorithm \ref{alg:PDifMP_boundedLambda} for state-dependent bounded jump rate functions. Throughout, we set the step size $h = 10^{-2}$. 

\paragraph{Summary statistics}

Due to the intrinsically stochastic and erratic nature of PDifMPs, simulated paths can vary substantially even when generated under identical parameter configurations. Nevertheless, the empirically observed ergodic behaviour of the PDifMP example problems enables the extraction of informative summary statistics from individual datasets derived from single sample paths, i.e. datasets $y_\theta=\{ x_\theta,N^j_\theta\}$.

To capture information about $\theta$ from the time-series component $x_\theta$ (which is itself influenced by the jump dynamics), we adopt the summary statistics originally proposed in \cite{Buckwar2020} and subsequently used in \cite{Ditlevsenetal2023,SAMSON2025108095} for ABC inference in SDE models. Specifically, we use $x_\theta$ to estimate both the underlying invariant density $f_{x_\theta}$ of the $X$-component of the process and its invariant spectral density $S_{x_\theta}$, via standard kernel density and periodogram estimators, respectively (cf. Appendix~\ref{appA.2}). 

Our analyses further indicate that including also the mean quadratic variation statistic
\begin{equation*}
    V_{x_\theta} = \frac{1}{N^x}\sum_{i=1}^{N^x} (x_{t_{i-1}} - x_{t_i})^2,
\end{equation*}
improves the estimation of the noise parameter $\sigma$.

To capture the hybrid nature of PDifMPs, which combine SDE and jump dynamics, we also include $N_\theta^j$ (the number of jumps generated under a given $\theta$) in the proposed set of summaries, without relying on the jump times. 

In summary, we define the PDifMP summary statistics of a dataset $y_\theta=\{ x_\theta,N^j_\theta\}$ as
\begin{equation}
\label{eq:summaries}
    s(y_\theta) := \{f_{x_\theta},S_{x_\theta},V_{x_\theta},N^j_{\theta}\}.
\end{equation}

\paragraph{Distance measures}

The proposed PDifMP summary statistics \eqref{eq:summaries} comprise both function-valued components ($f_{x_\theta}$ and $S_{x_\theta}$) and scalar components ($V_{x_\theta}$ and $N_\theta^j$). To quantify the discrepancy between two functions, we define $d_{\textrm{fun}}(\cdot,\cdot)$ as the sum of absolute pointwise differences. For scalar quantities, we use the absolute difference as the distance measure.

Combining these, we define the overall distance between two sets of summary statistics as
\begin{equation}
    d(s(y),s(y_\theta)) = w_1 d_{\textrm{fun}}(f_y,f_{y_\theta})  + w_2  d_{\textrm{fun}}(S_y,S_{y_\theta})  + w_3 |V_x - V_{x_\theta}| +  w_4 |N^j - N^j_{\theta}|,
    \label{eq:distance}
\end{equation}
where $s(\cdot)$ is as in \eqref{eq:summaries}, and $w_1,w_2,w_3,w_4$ are weights chosen to ensure that the four terms are of comparable magnitude. 

To compute these weights, one can, for instance, fix $w_1=1$ and generate multiple synthetic datasets $y_\theta$ under different parameter values $\theta$ drawn from the prior $\pi(\theta)$. Their corresponding distances to the observed dataset $y$ are then computed, and the respective weight is set to the reciprocal of the median of the resulting distances.

\subsection{Algorithm and implementation details}
\label{sec:implDet}

The basic ABC algorithm outlined at the beginning of Section \ref{sec:4:ABC} is computationally inefficient. This inefficiency arises because candidate parameter values $\theta^{'}$ are sampled from the prior distribution, which typically covers much larger parameter regions than the desired posterior distribution. As a result, a substantial amount of computational effort (such as model simulations, summary computations, and distance evaluations) is spent on candidate values $\theta^{'}$, which are ultimately rejected.

To overcome this limitation, sequential ABC algorithms have been developed, with the Sequential Monte Carlo ABC (SMC-ABC) algorithm being the current gold-standard method. In this sequential approach, one initially runs the basic ABC procedure (cf. the beginning of Section~\ref{sec:4:ABC}) until a certain number of parameter values have been accepted as samples from the approximate posterior. In this work, we use 100 accepted parameter values, referred to as particles. In subsequent iterations, new parameter candidates are obtained by resampling from the previously accepted values (according to their associated weights) and perturbing them with an appropriate perturbation kernel. We use multinomial resampling and the multivariate Gaussian perturbation kernel proposed in \cite{Filippi2013}, given by $\mathcal{N}(\theta',2\Sigma)$, where $\theta'$ denotes the resampled particle and $\Sigma$ is the weighted empirical covariance matrix of the accepted particles from the previous iteration.

The initial ABC threshold is set to $\delta=4$ and, in each subsequent iteration, updated to the median of the distances of the accepted particles from the previous iteration. Alternatively, the initial threshold may be selected from an ABC pilot run as a chosen percentile of the distances computed from trajectories simulated under parameter values sampled from the prior. The weights $w_1,w_2,w_3,w_4$ in the distance function~\eqref{eq:distance} are determined from such an ABC pilot run and set to the reciprocal of the median distance associated with each summary statistic.

This iterative procedure is repeated until a predefined stopping criterion is met, such as reaching a maximum number of synthetic data simulations (maximal computational budget) or a minimal percentage of accepted parameter values (minimal acceptance rate). In this work, we use the maximal computational budget as the stopping criterion. The optimal stopping criterion may vary depending on the model under consideration. Further details on the SMC-ABC algorithm can be found in \cite{Beaumont2009,DelMoral2012,Marin2012,Sisson2007}. A precise description of the specific algorithm used in this work is provided, e.g., in Algorithm 1 of \cite{SAMSON2025108095}. Package details for the implementation can be found in Appendix \ref{appA.2}.

We provide well-documented and user-friendly sample code, including complete algorithmic details and specifications, on GitHub. In particular, code for path simulation of the  PDifMP test problems (i.e., implementations of Algorithms \ref{alg:PathFlowPhi}--\ref{alg:PDifMP_boundedLambda} in Appendix \ref{appA.1}) is available at: \url{https://github.com/AgnesMall/Simulation_of_PDifMPs}. Furthermore, code for PDifMP inference via SMC-ABC can be found at: \url{https://github.com/AgnesMall/PDifMP_inference_SMC-ABC}. 


\section{Inference results}
\label{sec:5:Inference}

In this section, we apply the ABC method for inference in PDifMPs, as proposed in the previous section, to the set of test problems with empirically observed ergodic behaviour introduced in Section~\ref{sec:3:2:testProblems}. In Section~\ref{sec:5:1:Inference}, we examine the performance of the ABC algorithm for inferring $\theta$ as in \eqref{eq:theta} in test problems 1--4 under varying observation settings, time horizons, and constant jump rate functions $\Lambda \equiv \lambda$, and investigate the contribution of the individual summary statistics through an ablation study. In Section~\ref{sec:5:2:NonConstantLambda}, we extend this analysis to non-constant jump rate functions~$\Lambda$. Finally, in Section~\ref{sec:5.3_4parameter}, we evaluate the algorithm's performance when additional parameters (specifically those appearing in the drift coefficient of the underlying SDE; previously denoted by $\eta$) are included in the parameter vector $\theta$ \eqref{eq:theta}. Each inference experiment was repeated five times, but for clarity, we report only one representative trial. An analysis of the variability across the five runs is provided in Appendix \ref{app:C:MulipleRuns}.


\subsection{Inference under different observation settings}
\label{sec:5:1:Inference}

\paragraph{Inference in TP~1 (OU-PDifMP) for different parameter choices}

\begin{table}
    \centering
    \begin{tabular}{c!{\vrule width 1.2pt}c|c|c}
        Setting & $\sigma$ & $b$ & $\lambda$ \\ \noalign{\hrule height 1.2pt}
        1 & 1 & 2 & 0.1 \\ \noalign{\hrule height 1.2pt}
        2 & 1 & 2 & 0.2 \\ \noalign{\hrule height 1.2pt}
        3 & 2 & 4 & 0.2 
    \end{tabular} \\
    \caption{Parameter settings used to simulate the respective observed dataset for TP~1 (OU-PDifMP).}
    \label{tab:p6_settings}
\end{table}

\begin{figure}
    \centering
    \includegraphics[width=\textwidth]{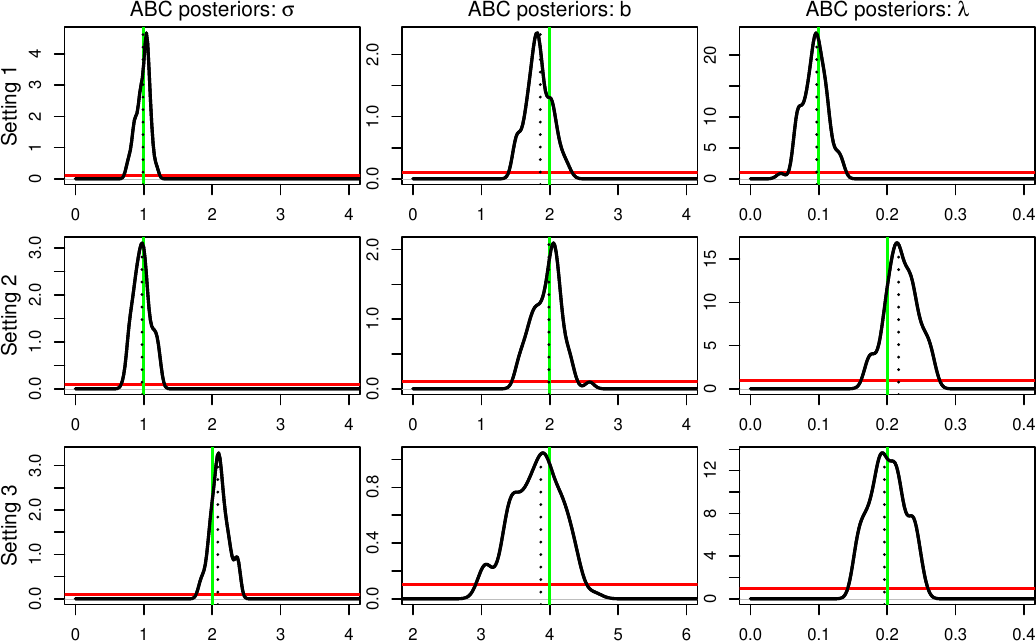}
    \caption{Marginal ABC posterior densities of the parameter vector $\theta$ \eqref{eq:theta} for TP~1 (OU-PDifMP), derived under three different observation settings.}
    \label{fig:p6_posterior1}
\end{figure}

For the first test problem, we estimate $\theta$ \eqref{eq:theta} based on three distinct simulated datasets. Each observed dataset is generated with Algorithm \ref{alg:PDifMP_constantLambda} in Appendix \ref{appA.1}, using a time horizon of $T=500$ and parameter values for $\sigma$, $b$, and $\lambda$ as specified in Table \ref{tab:p6_settings}. Setting~1 serves as the standard parameter configuration; in Setting~2, the jump rate of the process is increased, and in Setting~3, all parameters are increased relative to the standard~setting. 

Figure \ref{fig:p6_posterior1} reports the marginal ABC posterior densities (black solid lines) of $\theta$ \eqref{eq:theta}, obtained under the three different observed datasets. Each row corresponds to a specific observation setting, while each column represents a particular parameter. The posterior densities are compared with the corresponding prior densities (red horizontal lines, cf. \eqref{eq:priors}), for which only a subset of the domain is shown to improve interpretability. Across all settings and parameters, the posteriors are unimodal and cover the underlying true parameter values (green vertical lines). Moreover, the marginal ABC posterior medians (black dotted vertical lines) are consistently close to the true values. Note that the ABC algorithm was stopped after the computational budget (defined as the number of synthetic datasets simulated from the PDifMP) exceeded $10^4$.

\begin{figure}[t]
    \centering
    \includegraphics[width=\textwidth]{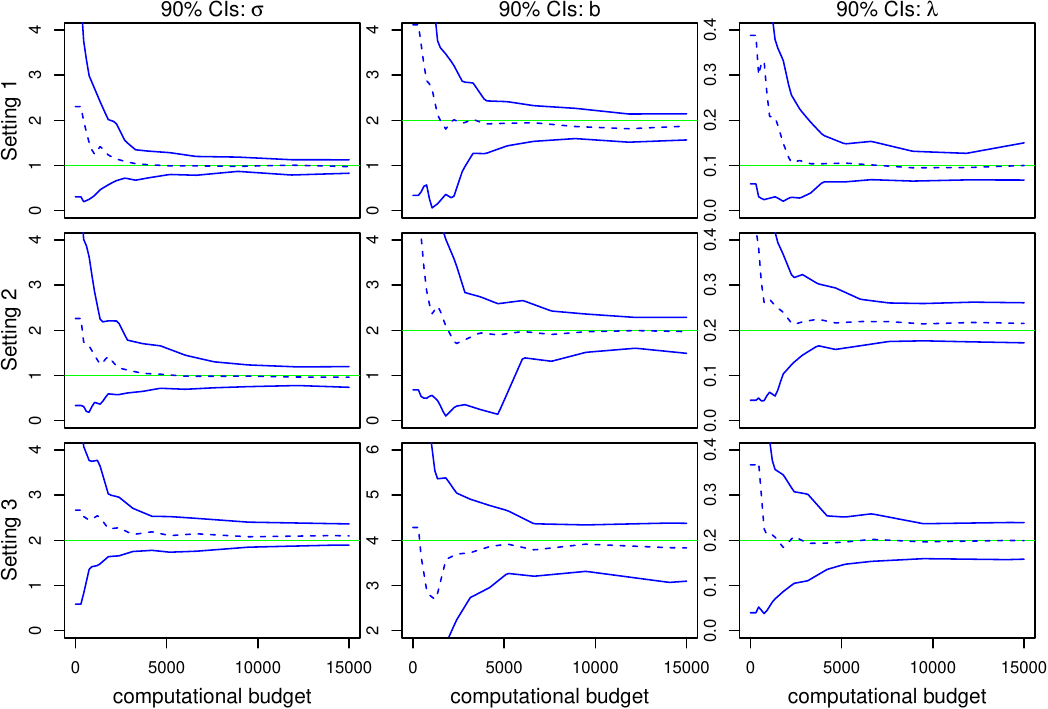}
    \caption{90\% CIs of the marginal posterior densities of the parameter vector $\theta$ \eqref{eq:theta} for TP~1 (OU-PDifMP) as functions of the computational budget, derived under three different observation settings.}
    \label{fig:p6_CI}
\end{figure}

Figure \ref{fig:p6_CI} shows the 90\% credible intervals (CIs) of the posteriors of $\theta$ \eqref{eq:theta} as functions of the computational budget, obtained under the three observation settings. As before, each row corresponds to a specific observation setting, and each column represents a particular parameter. Initially, the CIs (blue solid lines) span broad parameter regions influenced by the respective priors, but they become more narrow and approach the desired posterior regions as the computational budget increases. Correspondingly, the posterior medians (blue dashed lines) approach the underlying true parameter values (green horizontal lines) and approximate them well once the computational budget is sufficiently large. 

\begin{figure}[t]
    \centering
    \includegraphics[width=\textwidth]{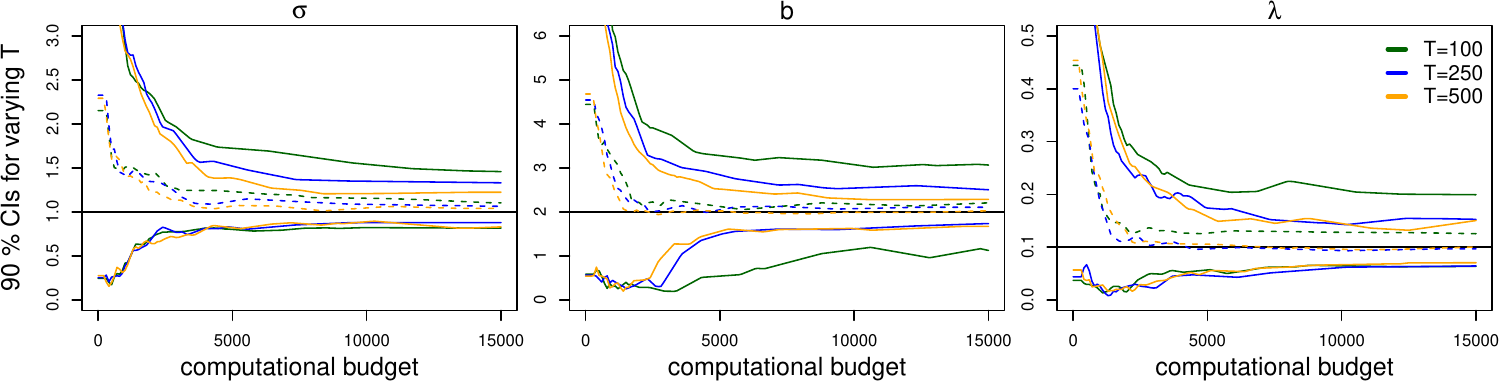}
    \caption{90\% CIs of the marginal posterior densities of the parameter vector $\theta$ \eqref{eq:theta} for TP~1 (OU-PDifMP), derived under three different time horizons $T$ for the standard parameter setting (Setting 1).}
    \label{fig:p6_T_CI}
\end{figure}

\paragraph{Inference in TP~1 (OU-PDifMP) for different time horizons}

\begin{figure}
    \centering
    \includegraphics[width=\textwidth]{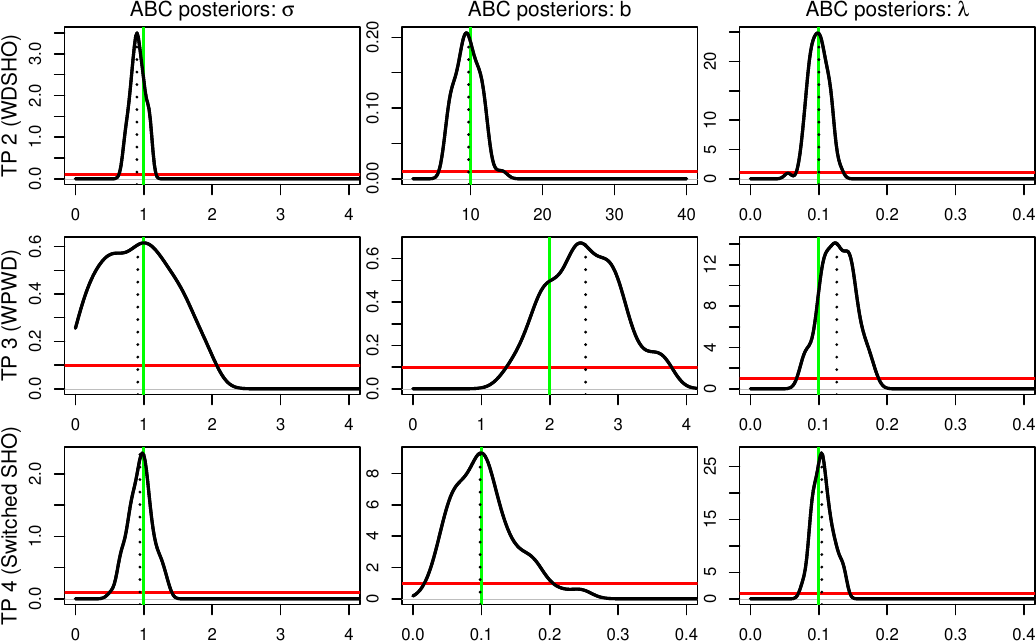}
    \caption{Marginal ABC posterior densities of the parameter vector $\theta$ \eqref{eq:theta} for TP~2 (WDSHO-PDifMP), TP~3 (WPWD-PDifMP), and TP~4 (SwitchedSHO-PDifMP).}
    \label{fig:posterior_2_4}
\end{figure}

\begin{figure}
    \centering
    \includegraphics[width=\textwidth]{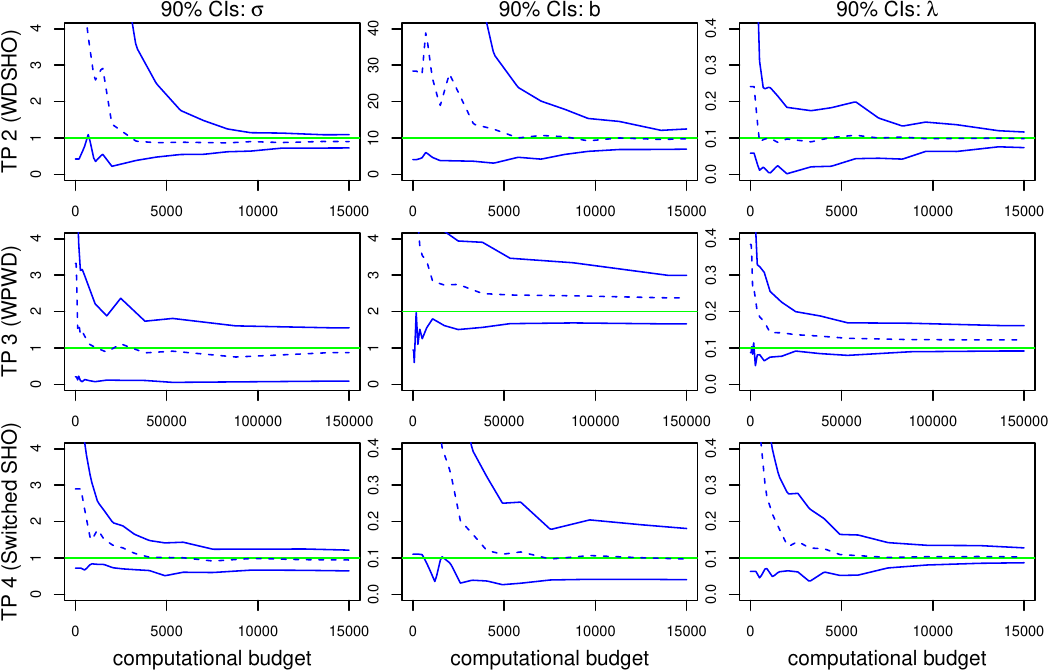}
    \caption{90\% CIs of the marginal posterior densities of the parameter vector $\theta$ \eqref{eq:theta} for TP~2 (WDSHO-PDifMP), TP~3 (WPWD-PDifMP), and TP~4 (SwitchedSHO-PDifMP), as functions of the computational budget.}
    \label{fig:CI_2_4}
\end{figure}

We now fix the standard parameter setting (Setting~$1$) and simulate observed datasets for three different time horizons $T$. Figure~\ref{fig:p6_T_CI} shows the corresponding $90\%$ CIs of the resulting posteriors for $\theta$ \eqref{eq:theta} as functions of the computational budget. Even for relatively small $T$, the posteriors cover the true parameter values. As the observation time horizon $T$ increases, the posteriors become narrower and the posterior medians (dashed lines) more closely approximate the true parameter values (horizontal lines). However, observing data over very large time horizons requires substantially greater computational effort, while yielding only marginal improvements in inference accuracy. A more detailed analysis of the impact of the time horizon on the inferential results for all test problems is provided in Appendix~\ref{app:B:timeHorizon}.

\paragraph{Inference in TP~2--4}

After validating the ABC method on the first test problem, we proceed to analyze the remaining ones. For each test problem, we consider a single observation setting, with the corresponding true parameter values listed in Table \ref{tab:p7p2p12_settings}. These settings are similar to the standard parameter setting (Setting 1) used for TP~1 (OU-PDifMP), except that the parameter $b$ is adjusted to ensure that TP~2 (WDSHO-PDifMP) and TP~4 (SwitchedSHO-PDifMP) are well defined. The time horizon $T$ is set to $1000$ for TP~2 (WDSHO-PDifMP), $1000$ for TP~3 (WPWD-PDifMP) and $5000$ for TP~4 (SwitchedSHO-PDifMP).  We also verified that the proposed ABC method yields similar results under alternative parameter settings. 

\begin{table}[t]
    \centering
    \begin{tabular}{c!{\vrule width 1.2pt}c|c|c}
        Test problem & $\sigma$ & $b$ & $\lambda$ \\ \noalign{\hrule height 1.2pt}
        TP~2 & 1 & 10 & 0.1 \\ \noalign{\hrule height 1.2pt}
        TP~3 & 1 & 2 & 0.1 \\ \noalign{\hrule height 1.2pt}
        TP~4 & 1 & 0.1 & 0.1 
    \end{tabular} \\
    \caption{Parameter settings used to simulate the respective observed dataset for TP~2 (WDSHO-PDifMP), TP~3 (WPWD-PDifMP), and TP~4 (SwitchedSHO-PDifMP).}
    \label{tab:p7p2p12_settings}
\end{table}

Figure \ref{fig:posterior_2_4} shows the marginal posterior densities (black solid lines) of $\theta$ \eqref{eq:theta} for test problems 2--4, along with the corresponding prior densities (red horizontal lines). As before, only a subset of the prior domain is displayed (cf. \eqref{eq:priors}). For all test problems, the posteriors are unimodal and cover the true parameter values (green horizontal lines), which are well approximated by the posterior medians (black dotted vertical lines). It can be observed that the posterior dispersions differ across parameters within the same test problem. For example, in TP~4 (SwitchedSHO-PDifMP), the posteriors of $\sigma$ and $\lambda$ have a smaller coefficient of variation than the posterior of $b$. This indicates that, in this test problem, the parameter $b$ can vary more than the others while still producing paths with comparable summaries. The algorithm was stopped after the computational budget exceeded $1.3 \times 10^4$ for TP~2 (WDSHO-PDifMP), $5\times 10^4$ for TP~3 (WPWD-PDifMP), and $1.3 \times 10^4$ for TP~4 (SwitchedSHO-PDifMP).

Figure \ref{fig:CI_2_4} shows the 90\% CIs of the posteriors of $\theta$ \eqref{eq:theta} for test problems 2--4 as functions of the computational budget. The x-axis for each parameter differs across the test problems, as for some test problems the posteriors converge to the desired region  faster than for others. As observed for TP~1 (OU-PDifMP), the CIs (blue solid lines) initially span broad parameter regions influenced by the respective priors, but they become narrower and converge toward the posterior regions as the computational budget increases. Correspondingly, the posterior medians (blue dashed lines) closely approximate the underlying true parameter values (green horizontal lines) once the computational budget is sufficiently large.

Finally, we note that for TP~3 (WPWD-PDifMP), the posteriors appear slightly less precise than for the other test problems. In particular, the lower bound of the 90\% CI for the parameter $\sigma$ remains close to the lower bound of the corresponding prior distribution throughout. Moreover, the posteriors for $b$ and $\lambda$ are not very well centred on the respective true values. As discussed in Section~\ref{sec:3:2:testProblems}, this may be related to the less pronounced agreement between space and time averages observed for TP~3, which may result from an insufficient time horizon or reflect a lack of theoretical ergodicity. In line with the observations in Section~\ref{sec:3:2:testProblems}, we also found that increasing the jump rate $\lambda$ and the time horizon $T$, for which the agreement between space and time averages improves (cf. Figure \ref{fig:KL_distance} and Table \ref{fig:tp3_full_analysis}), leads to narrower posterior distributions that are more closely centred around the true parameter values (figures not shown). In Section~\ref{sec:6:TP3}, we investigate TP~3 (WPWD-PDifMP) in more detail, examining how these space-time discrepancies propagate into the summary statistics and consequently affect the ABC inference. We also introduce an alternative set of summary statistics tailored to TP~3 that does not rely on ergodic assumptions, but requires the individual jump times to be observed.

\paragraph{Ablation study of the summary statistics for TP~1--4}
The inference results presented above were obtained using the full set of summary statistics \eqref{eq:summaries}. We now investigate the contribution of the individual summary statistics by repeating the inference with progressively extended summary sets. This also allows us to assess why the invariant density and invariant spectral density, which have previously been successfully used for parameter inference in SDEs \cite{Buckwar2020,Ditlevsenetal2023,SAMSON2025108095}, need to be extended by additional summaries in the PDifMP setting.

Specifically, we repeat the inference for TP1--TP4 using the same setup as above, varying only the set of summary statistics. We compare the resulting 90\% credible intervals of the marginal posteriors for $\theta$ \eqref{eq:theta} across the computational budget for three summary sets:
\begin{itemize}
\item $f + S$: the invariant density ($f$) and the invariant spectral density ($S$),
\item $f + S + N$: additionally including the number of jumps ($N$),
\item $f + S + N + V$: additionally including the mean quadratic variation statistic ($V$).
\end{itemize}

\begin{figure}
    \centering
    \includegraphics[width=1.0\textwidth]{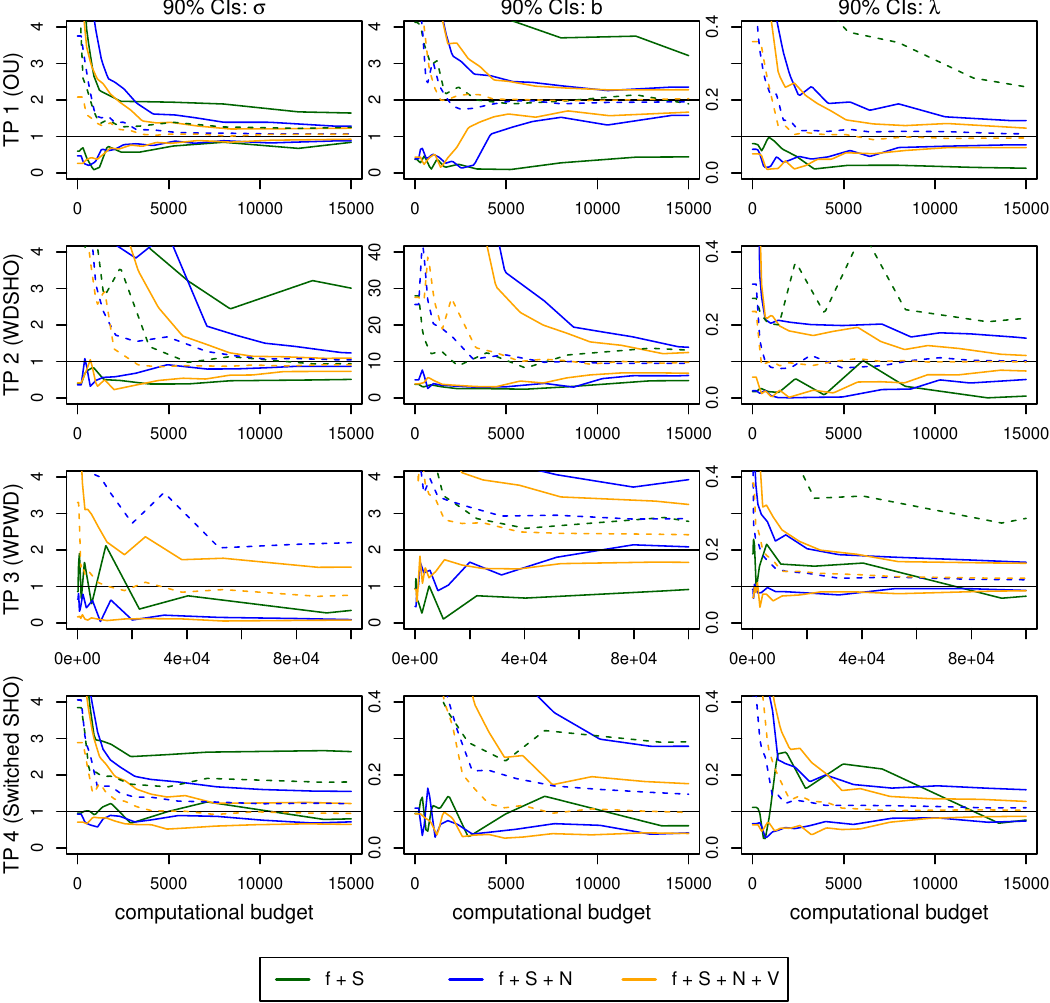}
    \caption{90\% CIs of the marginal posterior densities of the parameter vector $\theta$ \eqref{eq:theta} for all four test problems, with different sets of summaries for the standard parameter settings.}
    \label{fig:summary_ablation}
\end{figure}

As shown in Figure~\ref{fig:summary_ablation} (right column), using only $f+S$ does not allow the jump rate $\lambda$ to be identified, with the corresponding credible intervals (green solid lines) remaining very broad and the posterior medians (green dashed lines) lying far from the true parameter values (black horizontal lines). Note that the upper bounds of the credible intervals are even outside the displayed range in all cases. Including the number of jumps ($N$) allows $\lambda$ to be identified and also improves the estimation of the remaining parameters (see the blue lines). Including the mean quadratic variation statistic ($V$) further improves the inference for all parameters and test problems (see the orange lines), with a particularly important contribution to the estimation of $\sigma$ in TP3 (WPWD-PDifMP), for which the credible interval obtained with only $f+S+N$ does not narrow around the true value as the computational budget increases. Across all test problems and parameters, the full summary set $f+S+N+V$ provides the best estimates overall, with generally narrower credible intervals that are more closely centred around the true parameter values. 

\subsection{Inference under process-dependent jump rate functions}
\label{sec:5:2:NonConstantLambda}

In the previous experiments, the jump rate function was held constant, i.e. $\Lambda \equiv \lambda>0$. In this section, we extend our analysis to jump rate functions $\Lambda(X_t;\lambda)$, which depend on the state of the $X$-component of the PDifMP and a parameter $\lambda>0$. In particular, we will illustrate that successful inference of $\theta$ \eqref{eq:theta} via the proposed ABC method is also possible when $\lambda \in \theta$ appears as a parameter in $\Lambda$.    

\paragraph{Process-dependent jump rate functions}

The following jump rate functions, defined for $x \in D_1$, are investigated:
\begin{align}
    \text{sigmoid: }\Lambda(x;\lambda) & = \frac{\lambda}{1+e^{-x}}, \label{eq:sigmoid} \\
    \text{reducedCenter: }\Lambda(x;\lambda) & = \begin{cases}
        \lambda/2, & |x| \leq 2 \\
        \lambda, & \text{else}
    \end{cases}, \label{eq:reducedCenter} \\
    \text{cos: }\Lambda(x;\lambda) & = \lambda\cos(x) + \lambda. \label{eq:cos}
\end{align}

\begin{figure}[t]
    \centering
    \includegraphics[width=0.9\textwidth]{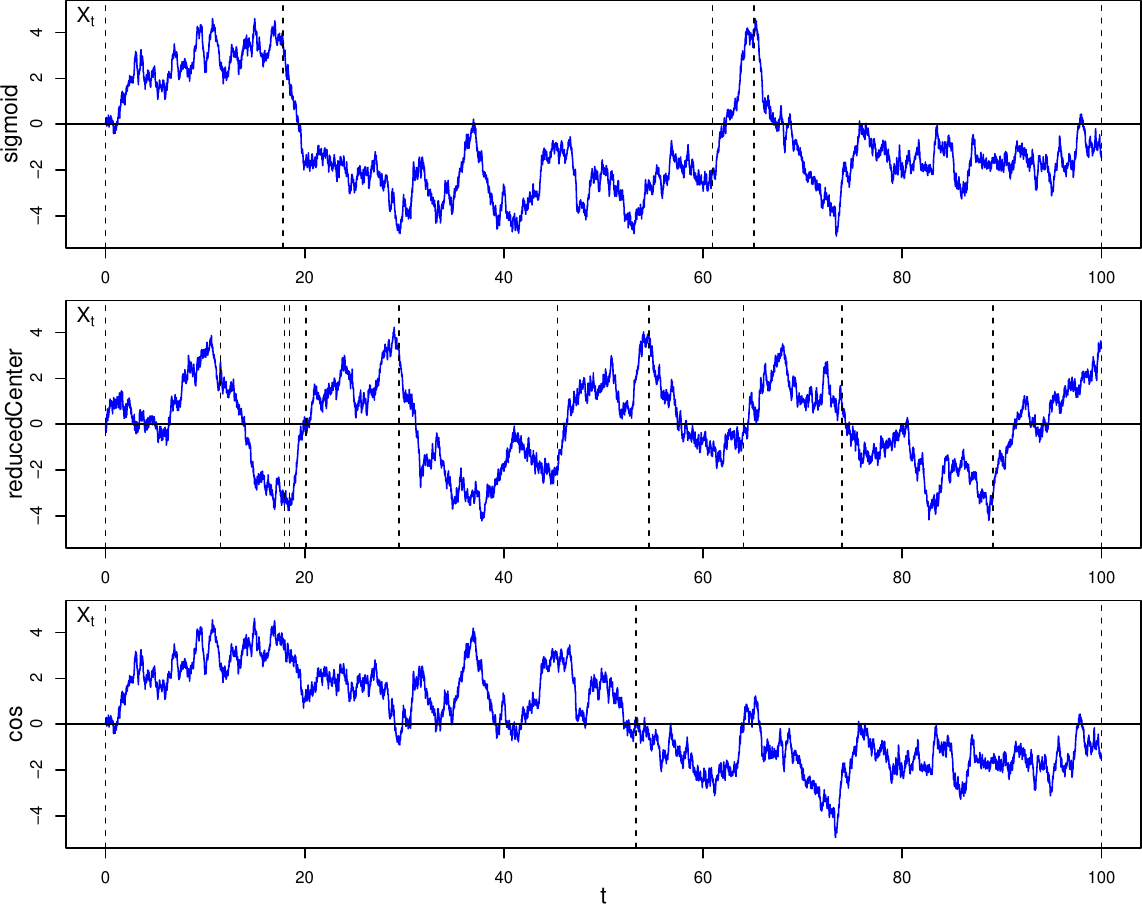}
    \caption{Paths of TP~1 (OU-PDifMP) with sigmoid, reducedCenter, and cos jump rate functions.}
    \label{fig:paths5_9_10}
\end{figure}

\begin{figure}[!htb]
    \centering
    \includegraphics[width=\textwidth]{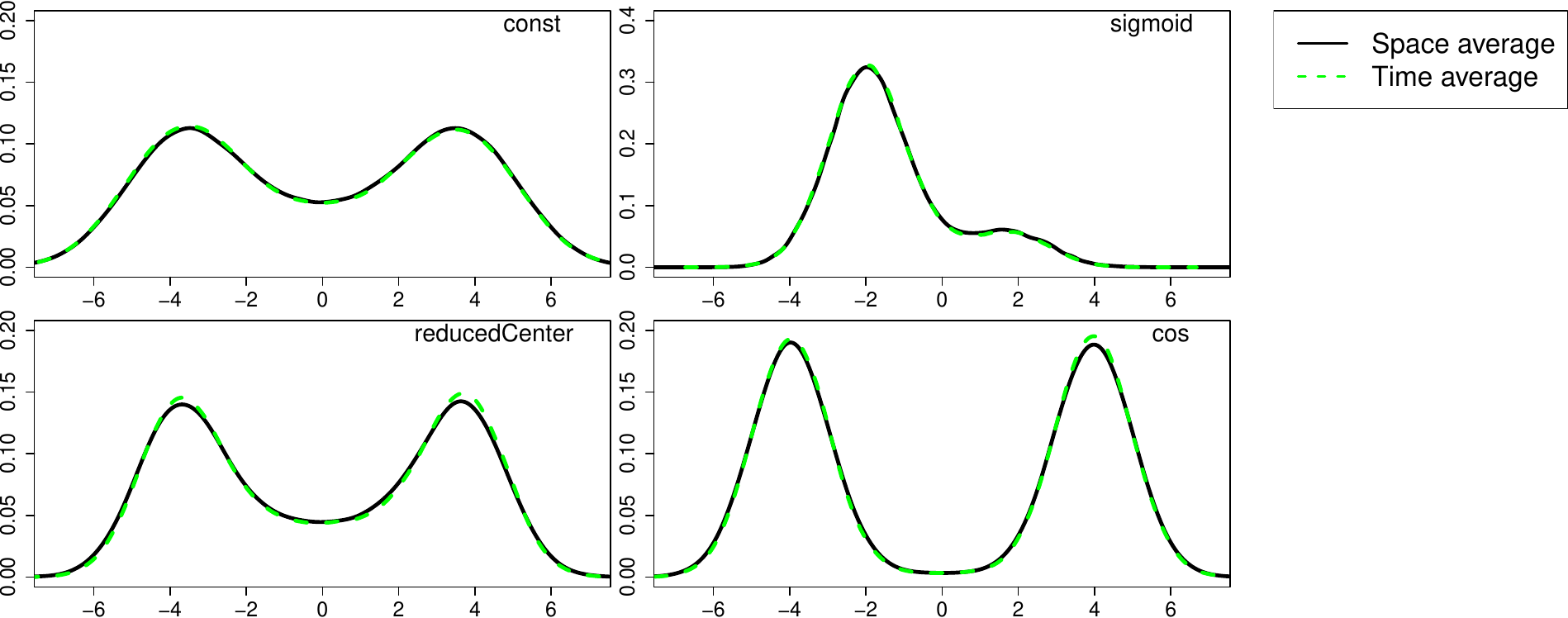}
    \caption{Space and time averages of TP~1 (OU-PDifMP) with constant, sigmoid, reducedCenter, and cos jump rate functions.}
    \label{fig:densities5_9_10}
\end{figure}

These jump rate functions represent different types of jumping behaviours and are bounded by a constant for all $x \in D_1$ so that the thinning method can be used (cf. Algorithm~\ref{alg:PDifMP_boundedLambda}). The sigmoid $\Lambda$ is monotonically increasing but bounded by $\lambda$. The reducedCenter $\Lambda$ is equal to $\lambda$ except for the region around 0, more precisely for $|x| \leq 2$, where it is reduced to $\lambda/2$ and overall again bounded by $\lambda$. The cos $\Lambda$ periodically increases and decreases, but is bounded by $2\lambda$. 

Figure \ref{fig:paths5_9_10} illustrates how the different jump rate functions affect the path behaviour of the OU-PDifMP. The sigmoid $\Lambda$ leads to fewer jumps for lower values of $X_t$, causing the path to move around the negative mean for a longer time than around the positive one. The reducedCenter $\Lambda$ produces fewer jumps when $X_t$ is close to the origin, making the PDifMP less likely to experience jumps during transitions of $X_t$ from one mean to the other. In contrast, the cos $\Lambda$ yields a lower jump intensity around $|X_t| = \pi$, which is close the mean in the chosen setting, and therefore the PDifMP is less likely to jump when the path is near the corresponding means.

Similar to Figure \ref{fig:space_time}, Figure \ref{fig:densities5_9_10} illustrates that densities derived from single paths (green dashed lines) of the OU-PDifMP with process-dependent jump rate functions closely resemble the densities computed from multiple realisations of $X_{t^{*}}$ for sufficiently large $t^{*}$ (black solid lines). The top left panel corresponds to the top left panel of Figure \ref{fig:space_time}, where the jump rate function is constant. The density of the PDifMP with the sigmoid jump rate function exhibits a bimodal distribution with an asymmetric profile, characterised by a dominant left mode and a smaller secondary mode on the right. For the reducedCenter jump rate function, some probability mass appears near the centre, although the two peaks remain clearly distinct. In contrast, the density obtained under the cos jump rate function shows two sharply separated peaks with minimal probability mass in the central region. Note that analogous figures can be obtained for the other test problems.

\begin{figure}[t]
    \centering
    \includegraphics[width=\textwidth]{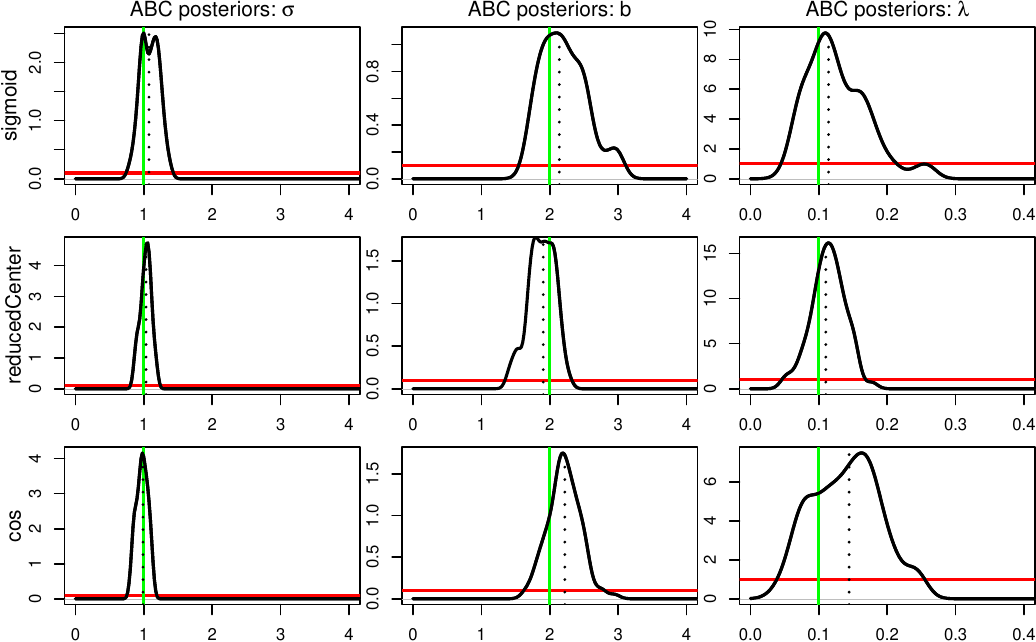}
    \caption{Marginal ABC posterior densities of $\theta$ \eqref{eq:theta} for TP~1 (OU-PDifMP) with sigmoid, reducedCenter, and cos jump rate functions.}
    \label{fig:posterior5_9_10}
\end{figure}

\paragraph{Inferential results}

We now repeat the inference of $\theta$ \eqref{eq:theta} reported in Section \ref{sec:5:1:Inference}, this time replacing the constant jump rate function $\Lambda \equiv \lambda$ with the process-dependent jump rate functions \eqref{eq:sigmoid}--\eqref{eq:cos}. We start with TP~1 (OU-PDifMP), focusing on Setting 1 from before with the same time horizon. 

Similar to before, Figure \ref{fig:posterior5_9_10} shows the marginal posterior densities (black solid lines) obtained under the three jump rate functions, compared to the prior densities (red horizontal lines, cf. \eqref{eq:priors}). The algorithm was stopped after the computational budget exceeded $4 \times 10^{4}$ for the sigmoid $\Lambda$, $10^{4}$ for the reducedCenter $\Lambda$, and $1.5 \times 10^{4}$ for the cos $\Lambda$. The corresponding 90\% CIs are reported in Figure \ref{fig:CI5_9_10}. The green solid lines indicate the true parameter values, while the dotted (respectively dashed) lines represent the posterior medians. Comparing these results with those from Setting 1 in Figure \ref{fig:p6_posterior1} and Figure \ref{fig:p6_CI}, the following observations can be made: 

\begin{figure}[t]
    \centering
    \includegraphics[width=\textwidth]{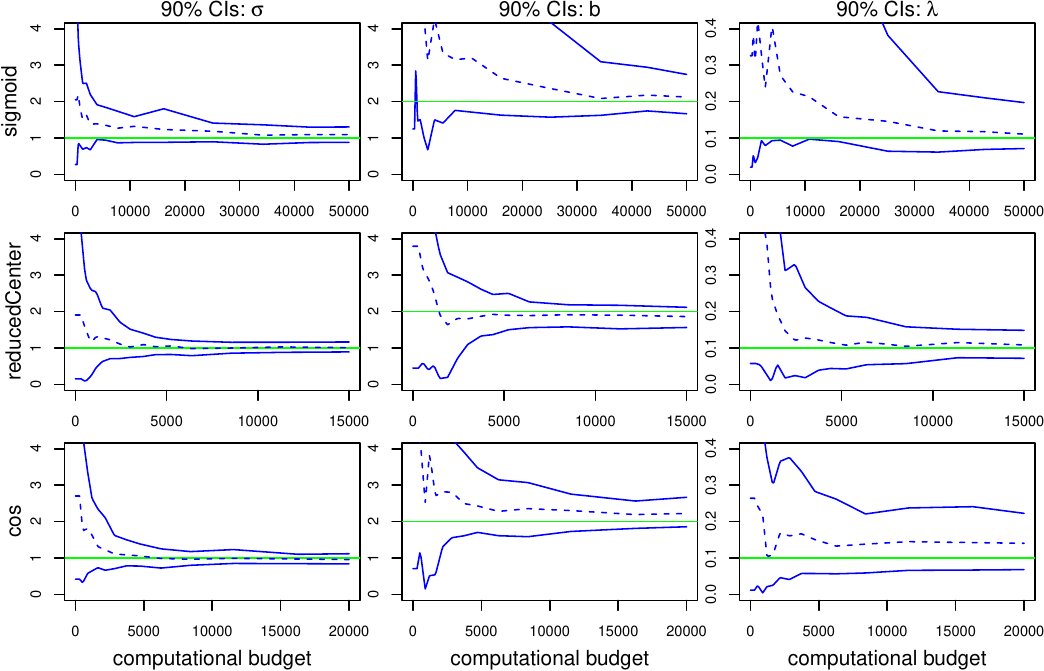}
    \caption{90\% CIs as functions of the computational budget of the marginal posterior densities of $\theta$ \eqref{eq:theta} for TP~1 (OU-PDifMP) with sigmoid, reducedCenter, and cos jump rate functions.}
    \label{fig:CI5_9_10}
\end{figure}

 The reducedCenter $\Lambda$ yields results that are very similar to those obtained with the constant jump rate function. Moreover, for all considered non-constant jump rate  functions, the inference of $\sigma$ closely matches that under the constant jump-rate case, indicating that the estimation of this parameter is highly robust to variations in the jump-rate dynamics. For the cos $\Lambda$, the inference of $b$ is likewise consistent with that under a constant $\Lambda$. In the case of the sigmoid $\Lambda$, the 90\% CIs for $b$ suggest a slower convergence to the target posterior region, although the final posterior estimate remains largely unaffected. Regarding the parameter $\lambda$, for both the cosine and sigmoid $\Lambda$ we also observe a slower convergence to the target posterior regime as well as broader CIs and posterior distributions with larger variances, compared to the constant $\Lambda$ case. Overall, it is noteworthy that the inference performance remains strong across all these different jump rate regimes. 

Similar results are obtained for TP~2 (WDSHO-PDifMP) and TP~4 (SwitchedSHO-PDifMP). For instance, the inference of the parameter $\sigma$ remains relatively robust across different choices of $\Lambda$. For the parameters $b$ and $\lambda$, we observe slower convergence to the target posterior region in certain cases, with final posterior estimates that are either largely unaffected or show slightly greater dispersion, compared to the constant $\Lambda$ case. For TP~3 (WPWD-PDifMP), some deviations are observed, as discussed in Section~\ref{sec:6:3:processDependentLambda}.


\subsection{Inferring additional parameters}
\label{sec:5.3_4parameter}

\begin{table}[t]
    \centering
    \begin{tabular}{c!{\vrule width 1.2pt}c|c}
        Test problem & $\eta$ & $\pi(\eta)$  \\ \noalign{\hrule height 1.2pt}
        TP~1 & 1 & $U(0,10)$ \\ \noalign{\hrule height 1.2pt}
        TP~4 & 20 & $U(2,100)$
    \end{tabular} \\
    \caption{True values and priors for $\eta$ in TP~1 (OU-PDifMP) and TP~4 (SwitchedSHO-PDifMP).}
    \label{tab:p6_12_setting_priors}
\end{table}

In the previous experiments, the drift parameter $\eta$ of the underlying SDE was fixed in all relevant PDifMP test problems. Here, we demonstrate that the extended parameter vector
\begin{equation}
\label{eq:theta_extended}
    \theta=(\sigma,b,\lambda,\eta),
\end{equation}
can also be successfully inferred with the proposed ABC method. In particular, we use again $\Lambda \equiv \lambda>0$ and focus on TP~1 (OU-PDifMP) and TP~4 (SwitchedSHO-PDifMP), considering the true values and prior distributions for $\eta$ reported in Table \ref{tab:p6_12_setting_priors}. In TP~1, the parameter $\eta$ represents the mean-reversion rate, while in TP~4 it influences the oscillation frequencies. The other parameters follow the standard setting, and the time horizon remains the same for each test problem.

\begin{figure}[t]
    \centering
    \includegraphics[width=\textwidth]{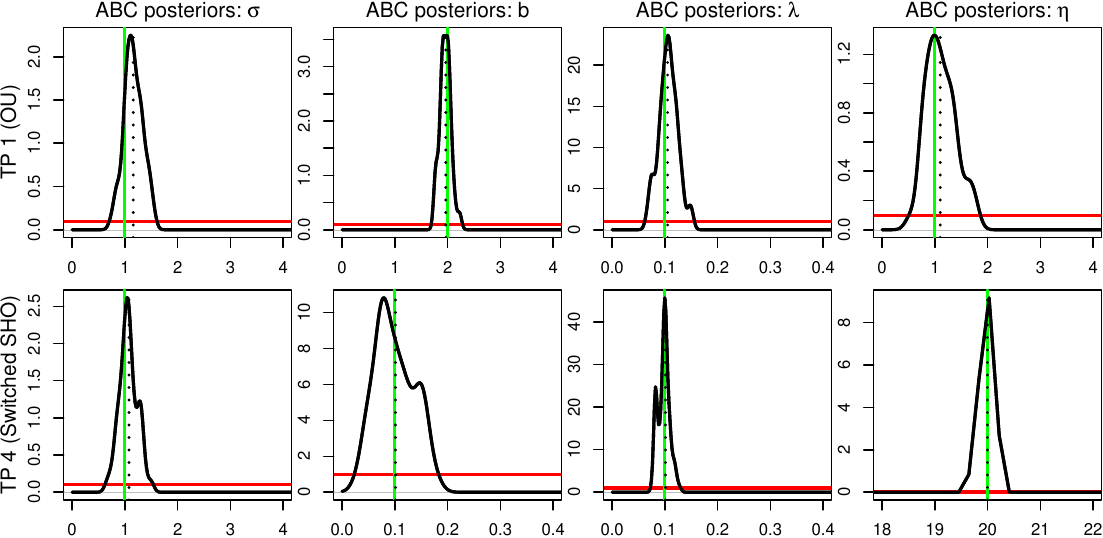}
    \caption{Marginal ABC posterior densities of the parameter vector $\theta$ \eqref{eq:theta_extended} for TP~1 (OU-PDifMP) and TP~4 (SwitchedSHO-PDifMP).}
    \label{fig:posterior_6_12}
\end{figure}

\begin{figure}[!htbp]
    \centering
    \includegraphics[width=\textwidth]{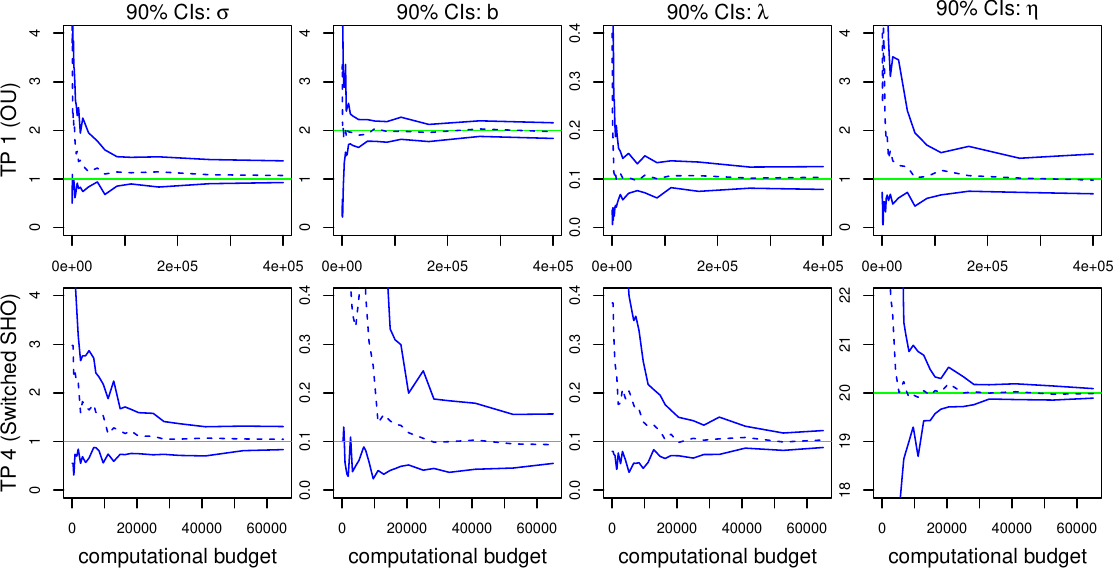}
    \caption{90\% CIs of the marginal posterior densities of the parameter vector $\theta$ \eqref{eq:theta_extended} for TP~1 (OU-PDifMP) and TP~4 (SwitchedSHO-PDifMP), as functions of the computational budget.}
    \label{fig:CI_6_12}
\end{figure}

Figures \ref{fig:posterior_6_12} and \ref{fig:CI_6_12} show the corresponding marginal posterior densities and 90\% CIs, respectively. Comparing these results with those from Setting~1 in Figures \ref{fig:p6_posterior1} and \ref{fig:p6_CI} (for TP~1), and with those from the bottom panels of Figures \ref{fig:posterior_2_4} and \ref{fig:CI_2_4} (for TP~4), it becomes evident that the inference of $\sigma$, $b$, and $\lambda$ is not degraded by adding the additional parameter $\eta$. Moreover, in both test problems, $\eta$ is estimated accurately. In particular, for TP~4 (SwitchedSHO-PDifMP), the marginal posterior of $\eta$ is  well centred on the true value and exhibits relatively low variance (see the bottom right panels of \ref{fig:posterior_6_12} and \ref{fig:CI_6_12}). The algorithm was stopped after the computational budget exceeded $1.5 \times 10^5$ for TP~1 (OU) and $5 \times 10^{4}$ for TP~4 (Switched SHO). Since an additional parameter $\eta$ is inferred, the convergence is slower, and therefore a larger computational budget was used compared to the settings with three parameters.

\section{Deeper investigation of TP~3 (WPWD-PDifMP)}
\label{sec:6:TP3}

In this section, we present additional analyses for TP~3 (WPWD-PDifMP). As discussed in Section~\ref{sec:3:2:testProblems}, TP~3 exhibits distinct behaviour compared with the other test problems when comparing space and time averages. The observed discrepancies may be related to a lack of theoretical ergodicity or to the considerably longer time horizons required to obtain small discrepancies between space and time averages. This behaviour is also apparent in the ABC inference results presented in Section~\ref{sec:5:1:Inference}, where the posterior distributions are wider and less closely centred around the true parameter values than for the other test problems (cf. Figures \ref{fig:posterior_2_4} and \ref{fig:CI_2_4}). We therefore investigate how the observed space-time discrepancies affect the summary statistics associated with the stationary distribution and, consequently, the ABC inference results.

\begin{figure}[t]
    \centering
    \includegraphics[width=0.8\textwidth]{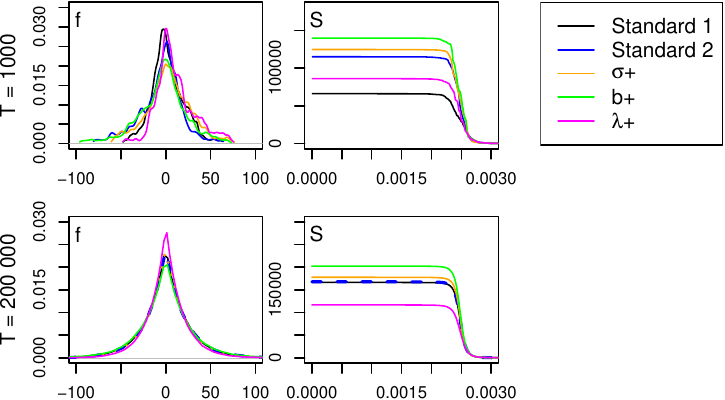}
    \caption{Density and spectral density for TP~3 (WPWD-PDifMP) under different parameter settings for the time horizon $T = 1000$ and $T = 200 000$.}
    \label{fig:summaries_TP3}
\end{figure}

Figure~\ref{fig:summaries_TP3} shows the density and spectral density summary statistics for TP~3 (WPWD-PDifMP) for the standard parameter setting (represented in black and blue) together with three additional settings in which one of the parameter values was increased. For the ABC algorithm to distinguish between parameter values, summaries generated under the same parameter setting should be similar, whereas summaries generated under different parameter settings should be sufficiently distinct. For $T = 1000$ (top panels), the spectral densities obtained under the same parameter setting (black and blue curves) show noticeable differences, and similar discrepancies are observed for the empirical densities. These differences decrease as the time horizon is increased to $T=200000$ (bottom panels). The standard parameter setting uses the relatively small jump rate $\lambda=0.1$. Based on the analysis in Section~\ref{sec:3:2:testProblems}, for larger values of $\lambda$, a shorter time horizon than $T=200000$ is expected to be sufficient to obtain a similar agreement between summaries generated under the same parameter setting, although the required time horizons may still be relatively large. At the same time, the differences between summaries generated under different parameter settings can be comparatively small for $T=1000$. Consequently, sufficiently long time horizons are important for these summary statistics to provide reliable information for ABC inference.

Since simulating sufficiently long trajectories for reliable inference based on the invariant density and invariant spectral density may be computationally infeasible, we additionally consider summary statistics that do not rely on ergodic assumptions and can make use of information available over shorter time horizons. Section~\ref{sec:6:1:TP3_extension} introduces a TP~3-specific set of summary statistics under the additional assumption that the individual jump times are observed and presents the corresponding ABC inference results. Furthermore, Section~\ref{sec:6:3:processDependentLambda} provides an extension to non-constant jump rate functions~$\Lambda$.

\subsection{Observation of jump times and adaptation of summary statistics}
\label{sec:6:1:TP3_extension}

In all previous experiments, we assumed that the observed data were of the form \eqref{eq:observed_y}. In addition to the number of jumps $N^j$, we now assume to observe the exact times of these $N^j$ jumps, i.e. the observed dataset \eqref{eq:observed_y} extends to
\vspace{-0.2cm}
\begin{equation}\label{eq:observed_y_extended}
    y=\{ x,j \}, \quad \text{where } j=(j_k)_{k=0}^{N^j-1}.
\end{equation}
This extension enables the construction of additional summary statistics, which may improve the inference of $\theta$~\eqref{eq:theta} using the investigated ABC method.

For instance, given a dataset $y_\theta=\{x_\theta,j_\theta\}$, we can define the summary statistic
\begin{equation*}
    \textrm{slope}_{y_\theta} = \text{median}\left\{\left|\frac{x_{j_{k+1}} - x_{j_k}}{j_{k+1} - j_{k}}\right|,  \quad k = 0,\dots, {N}^j_\theta-1\right\},
\end{equation*} 
which incorporates all components of $y_\theta$. This statistic corresponds to the median of the absolute slopes of the $X$-component between consecutive jumps $j_k$ and $j_{k+1}$, and is particularly informative for TP~3 (WPWD-PDifMP), where the parameter $b$ is related to the rate at which the time-series increases (or decreases) between jumps. Note that for this statistic we only consider jumps with moves (i.e., a change of the $Z$-component from $b$ to $-b$ or vice versa indeed happened). Together with the mean quadratic variation statistic and the number of jumps, we obtain the following TP3-specific set of summary statistics:
\begin{equation}
\label{eq:summaries_extended}
    s(y_\theta):=\{ V_{x_\theta},N_\theta^j, \textrm{slope}_{y_\theta} \}.
\end{equation}

\begin{figure}[!htbp]
    \centering
    \includegraphics[width=\textwidth]{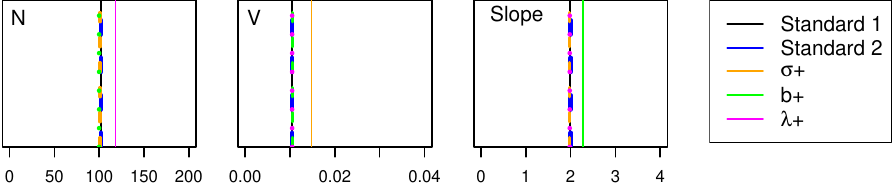}
    \caption{TP3-specific summary set (\ref{eq:summaries_extended}) under different parameter settings for the time horizon $T = 1000$.}
    \label{fig:summaries_TP3_3sum}
\end{figure}

Figure~\ref{fig:summaries_TP3_3sum} is analogous to Figure~\ref{fig:summaries_TP3}, but evaluates the alternative set of summary statistics \eqref{eq:summaries_extended}. The figure shows that changes in the individual parameters primarily affect different summary statistics, allowing the corresponding parameter settings to be distinguished.

\begin{figure}[t]
    \centering
    \includegraphics[width=\textwidth]{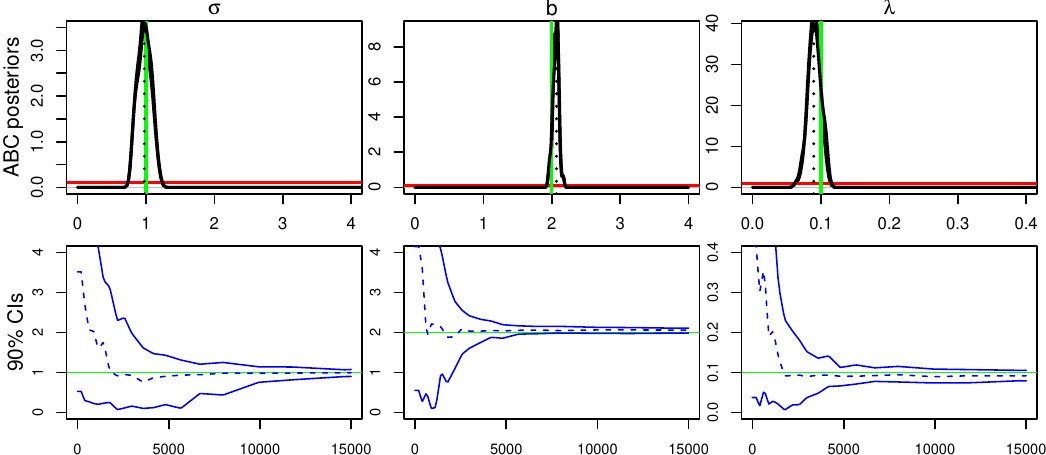}
    \caption{Marginal ABC posteriors densities of $\theta$ \eqref{eq:theta} for TP~3 (WPWD-PDifMP) and 90\% CIs as functions of the computational budget using the summary statistics \eqref{eq:summaries_extended}.}
    \label{fig:posterior_CI_2_S}
\end{figure}

We now repeat the inference of $\theta$ \eqref{eq:theta} for TP~3 (WPWD-PDifMP), as reported in Section~\ref{sec:5:1:Inference}, replacing the observed data \eqref{eq:observed_y} and general set of summary statistics \eqref{eq:summaries} with the observed data \eqref{eq:observed_y_extended} and the TP3-specific summary set \eqref{eq:summaries_extended}, respectively. In Figure \ref{fig:posterior_CI_2_S}, we present the corresponding estimation results. The top panels show the marginal posterior densities obtained when stopping the algorithm after the computational budget exceeded $10^4$, and the bottom panels display the respective 90\% CIs as functions of the computational budget. Using the TP3-specific summary set yields a clear improvement for all parameters, with narrower posterior distributions that are more closely centred around the true values (cf. Figures~\ref{fig:posterior_2_4} and~\ref{fig:CI_2_4}).

We emphasize, however, that the general set of summary statistics \eqref{eq:summaries} still provides reasonable inference results despite the more pronounced discrepancies between space and time averages observed for TP~3 and the uncertainty regarding its theoretical ergodicity.  Moreover, unlike the TP3-specific summary set \eqref{eq:summaries_extended}, the general set does not require the exact jump times to be observed, which may be an important advantage in practical applications. As discussed in Section~\ref{sec:3:2:testProblems}, the discrepancies between space and time averages for TP~3 are particularly pronounced for small jump rates. For larger jump rates, shorter time horizons may therefore be sufficient for inference based on the general summaries \eqref{eq:summaries}, although the required time horizons may still be relatively large. The TP3-specific summary set provides an alternative when sufficiently long trajectories are computationally infeasible and the individual jump times are available.

\subsection{Process-dependent jump rate functions}
\label{sec:6:3:processDependentLambda}

\begin{figure}[t]
    \centering
    \includegraphics[width=\textwidth]{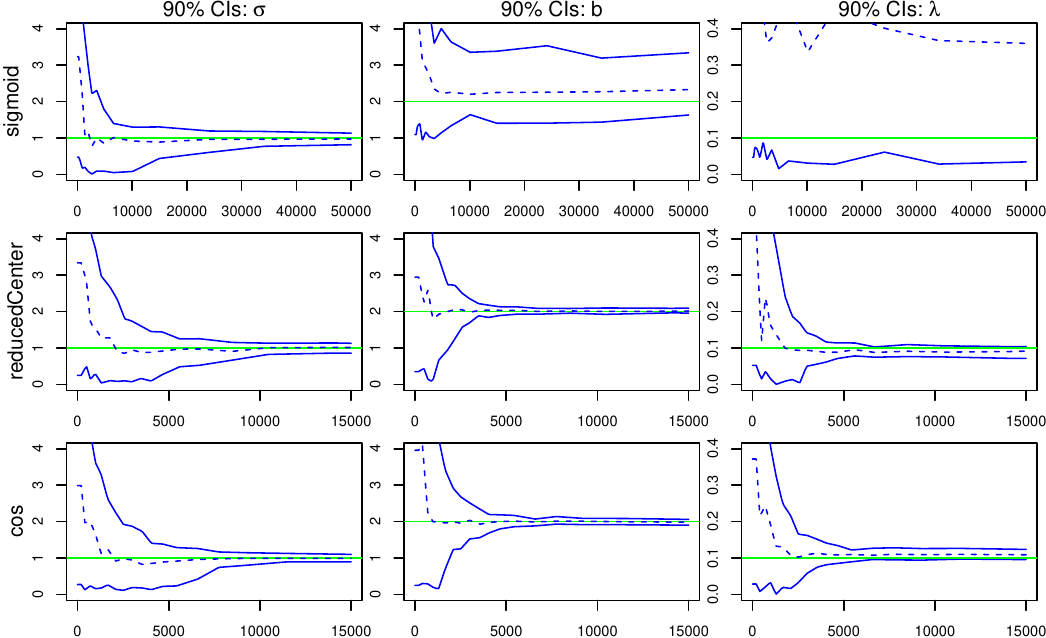}
    \caption{90\% CIs as functions of the computational budget of the marginal posterior densities of $\theta$ \eqref{eq:theta} for TP~3 (WPWD-PDifMP) with sigmoid, reducedCenter, and cos jump rate functions. }
    \label{fig:CI_18_19_22}
\end{figure}

\begin{figure}[htbp!]
    \centering
    \includegraphics[width=0.93\textwidth]{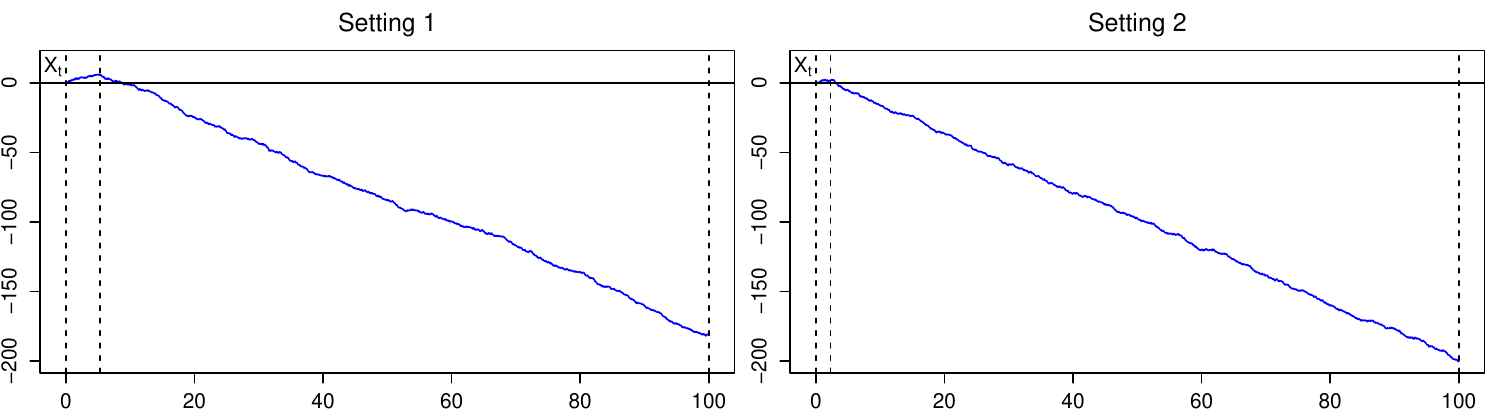}
    \caption{Paths of TP~3 (WPWD-PDifMP) with sigmoid jump rate function under two different parameter settings.}
    \label{fig:tp22_paths}
\end{figure}

In this section, we analyse how state-dependent jump rate functions $\Lambda$ affect the inferential results for TP~3 (WPWD-PDifMP). Specifically, we consider the jump rate functions \eqref{eq:sigmoid}, \eqref{eq:reducedCenter}, and \eqref{eq:cos}, respectively, and estimate $\theta$ \eqref{eq:theta}, where the parameter $\lambda$ now appears within these functions. The time horizon is again set to $1000$, as in the previous experiments for TP~3, and the summaries \eqref{eq:summaries_extended} are used. Figure \ref{fig:CI_18_19_22} shows the corresponding 90\% CIs for each of the three inference~experiments.

The reducedCenter and cos rate functions yield inference results that are similarly accurate to those obtained in the constant $\lambda$ case (cf. the bottom panels of Figure~\ref{fig:posterior_CI_2_S}). In contrast, under the sigmoid $\Lambda$ (top panels), the parameter $\lambda$ cannot be reliably estimated. This occurs because the jump rate decreases for smaller state values, causing sample paths to drift towards minus infinity and eventually preventing further jumps to happen, even for large values of $\lambda$. Figure~\ref{fig:tp22_paths} illustrates two representative sample paths for this setup, comparing the baseline configuration (Setting 1 with $\lambda=0.1$) with Setting 2 ($\lambda=0.9$). Both trajectories exhibit only a single jump over the considered time horizon, making it impossible to infer the jump rate from the jump count alone. Consequently, the posterior uncertainty for the parameter $b$ is larger, as the small number of jumps provides fewer trajectory segments from which to compute the slope summary. In contrast, the parameter $\sigma$ continues to be estimated accurately, illustrating that the TP3-specific summary statistics retain useful information about $\sigma$ even in this degenerate regime.


\section{Conclusion and future work}
\label{sec:7:conclusion}

In this manuscript, we have adapted the approximate Bayesian computation (ABC) methodology for parameter inference in piecewise diffusion Markov process (PDifMP) models that empirically exhibit ergodic behaviour. Specifically, we provided detailed algorithms for simulating PDifMP sample paths and embedded them within the ABC framework. Furthermore, we extended the summary statistics proposed in \cite{Buckwar2020} for stochastic differential equation (SDE) models to account for the hybrid characteristics of PDifMPs, achieving particular effectiveness in ergodic systems.  

The proposed ABC approach successfully inferred model parameters across a range of representative PDifMP test problems that empirically exhibited ergodic behaviour. It performed robustly well under various observation settings, including cases with constant and state-dependent jump rate functions, as well as in challenging scenarios where the jump and/or diffusion components were only partially observed. The ablation analysis further demonstrated the contribution of the individual summary statistics to the inference results. Our results also indicate that inference accuracy can be further improved by incorporating problem-specific summary statistics. Although this study primarily focused on estimating one parameter per characteristic model component, the method demonstrated strong performance even when additional parameters were included in the parameter vector to be inferred. Moreover, since the example PDifMPs considered here are generic and broadly applicable, our findings suggest that the proposed ABC framework holds promise for a wide range of stochastic hybrid models. Accompanying \texttt{R} code is provided to facilitate implementation and practical use of the method.

This work also opens several avenues for future research. First, while our empirical analyses provide strong indications of ergodicity in the considered test problems, a theoretical analysis of ergodicity remains an open problem. In particular, for TP~3 (WPWD-PDifMP), our results raise some doubts about whether this model is ergodic in general. Second, an important next step will be to extend the ABC methodology to non-ergodic PDifMPs, such as the model recently proposed in~\cite{buckwar2024americanoptionpricingusing}. Third, we aim to extend the ABC framework to more complex PDifMP models. The PDifMP test problems studied here involved exactly solvable SDEs, whereas many real-world applications will require efficient and reliable numerical methods to approximate the continuous component. Additionally, we considered state-dependent rate functions with natural upper bounds. However, in applications where such bounds are unavailable, constructing tight artificial bounds will be crucial to maintaining the efficiency of the underlying thinning algorithm. Finally, we plan to apply the developed inference framework to real-world data to further demonstrate its practical~potential.



\begin{thebibliography}{10}
	
	\bibitem{Beaumont2009}
	M.~A. Beaumont, J.-M. Cornuet, J.-M. Marin, and C.~P. Robert.
	\newblock {Adaptive approximate Bayesian computation}.
	\newblock {\em Biometrika}, 96(4):983--990, 2009.
	
	\bibitem{Beaumont2002}
	M.~A. Beaumont, W.~Zhang, and D.~J. Balding.
	\newblock Approximate {B}ayesian computation in population genetics.
	\newblock {\em Genetics}, 162(4):2025--2035, 2002.
	
	\bibitem{berg1972chemotaxis}
	H.~C. Berg and D.~A. Brown.
	\newblock Chemotaxis in {E}scherichia {C}oli analysed by three-dimensional
	tracking.
	\newblock {\em Nature}, 239(5374):500--504, 1972.
	
	\bibitem{blom1988piecewise}
	H.~A.~P. Blom.
	\newblock From piecewise deterministic to piecewise diffusion markov processes.
	\newblock In {\em Proceedings of the 27th IEEE Conference on Decision and
		Control}, pages 1978--1983. IEEE, 1988.
	
	\bibitem{buckwar2023stochastic}
	E.~Buckwar, M.~Conte, and A.~Meddah.
	\newblock A stochastic hierarchical model for low grade glioma evolution.
	\newblock {\em Journal of Mathematical Biology}, 86(6):89, 2023.
	
	\bibitem{buckwar2024americanoptionpricingusing}
	E.~Buckwar, S.~Desmettre, A.~Mallinger, and A.~Meddah.
	\newblock American option pricing using generalised stochastic hybrid systems.
	\newblock {\em Journal of Stochastic Analysis}, 6(1):Article 5, 2025.
	
	\bibitem{buckwar2025numerical}
	E.~Buckwar and A.~Meddah.
	\newblock {Numerical approximations and convergence analysis of piecewise
		diffusion Markov processes, with application to glioma cell migration}.
	\newblock {\em Applied Mathematics and Computation}, 491:129233, 2025.
	
	\bibitem{buckwar2011exact}
	E.~Buckwar and M.~G. Riedler.
	\newblock An exact stochastic hybrid model of excitable membranes including
	spatio-temporal evolution.
	\newblock {\em Journal of Mathematical Biology}, 63(6):1051--1093, 2011.
	
	\bibitem{Buckwar2020}
	E.~Buckwar, M.~Tamborrino, and I.~Tubikanec.
	\newblock {Spectral density-based and measure-preserving ABC for partially
		observed diffusion processes. An illustration on Hamiltonian SDEs}.
	\newblock {\em Statistics and Computing}, 30(3):627--648, 2020.
	
	\bibitem{bujorianu2003reachability}
	M.~L. Bujorianu and J.~Lygeros.
	\newblock Reachability questions in piecewise deterministic {M}arkov processes.
	\newblock In {\em International Workshop on Hybrid Systems: Computation and
		Control}, pages 126--140. Springer, 2003.
	
	\bibitem{Bujorianu2006}
	M.~L. Bujorianu and J.~Lygeros.
	\newblock {\em Toward a General Theory of Stochastic Hybrid Systems}, pages
	3--30.
	\newblock Springer Berlin Heidelberg, Berlin, Heidelberg, 2006.
	
	\bibitem{cassandras2018SHS}
	C.G. Cassandras and J.~Lygeros.
	\newblock {\em Stochastic Hybrid Systems}.
	\newblock Automation and Control Engineering. CRC Press, 2018.
	
	\bibitem{cloez2017probabilistic}
	B.~Cloez, R.~Dessalles, A.~Genadot, F.~Malrieu, A.~Marguet, and R.~Yvinec.
	\newblock Probabilistic and piecewise deterministic models in biology.
	\newblock {\em ESAIM: Proceedings and Surveys}, 60:225--245, 2017.
	
	\bibitem{Creel2015}
	M.~Creel and D.~Kristensen.
	\newblock {ABC of SV: Limited information likelihood inference in stochastic
		volatility jump-diffusion models}.
	\newblock {\em Journal of Empirical Finance}, 31:85--108, 2015.
	
	\bibitem{davis1984piecewise}
	M.~H.~A. Davis.
	\newblock {Piecewise-deterministic Markov processes: A general class of
		non-diffusion stochastic models}.
	\newblock {\em Journal of the Royal Statistical Society: Series B
		(Methodological)}, 46(3):353--376, 1984.
	
	\bibitem{DelMoral2012}
	P.~Del~Moral, A.~Doucet, and A.~Jasra.
	\newblock {An adaptive sequential Monte Carlo method for approximate Bayesian
		computation}.
	\newblock {\em Statistics and Computing}, 22:1009--1020, 2012.
	
	\bibitem{desmettre2025hybrid}
	S.~Desmettre, D.~Khurana, and A.~Meddah.
	\newblock The hybrid exact scheme for the simulation of first-passage times of
	jump-diffusions with time-dependent thresholds.
	\newblock {\em arXiv preprint arXiv:2511.00155}, 2025.
	
	\bibitem{DesmettreKhuranaMeddah2025}
	S.~Desmettre, D.~Khurana, and A.~Meddah.
	\newblock First-passage time for {PDifMPs}: An asymptotically exact simulation
	approach for time varying thresholds.
	\newblock {\em Journal of Scientific Computing}, 108(94), 2026.
	
	\bibitem{Ditlevsenetal2023}
	S.~Ditlevsen, M.~Tamborrino, and I.~Tubikanec.
	\newblock {Network inference via approximate Bayesian computation. Illustration
		on a stochastic multi-population neural mass model}.
	\newblock {\em Annals of Applied Statistics}, 19(4):3304--3329, 2025.
	
	\bibitem{Rcpp}
	D.~Eddelbuettel and R.~Fran\c{c}ois.
	\newblock {Rcpp}: Seamless {R} and {C++} integration.
	\newblock {\em Journal of Statistical Software}, 40(8):1--18, 2011.
	
	\bibitem{Filippi2013}
	S.~Filippi, C.~P. Barnes, J.~Cornebise, and M.~P.~H. Stumpf.
	\newblock {On optimality of kernels for approximate Bayesian computation using
		sequential Monte Carlo}.
	\newblock {\em Statistical Applications in Genetics and Molecular Biology},
	12(1):87--107, 2013.
	
	\bibitem{Frazier2019}
	D.~T. Frazier, W.~Maneesoonthorn, G.~M. Martin, and B.~P.~M. McCabe.
	\newblock {Approximate Bayesian forecasting}.
	\newblock {\em International Journal of Forecasting}, 35(2):521--539, 2019.
	
	\bibitem{guo2010piecewise}
	H.~Guo, A.~Mettas, G.~Sarakakis, and P.~Niu.
	\newblock Piecewise nhpp models with maximum likelihood estimation for
	repairable systems.
	\newblock In {\em 2010 Proceedings-Annual Reliability and Maintainability
		Symposium (RAMS)}, pages 1--7. IEEE, 2010.
	
	\bibitem{ishijima2011regime}
	H.~Ishijima and M.~Uchida.
	\newblock The regime switching portfolios.
	\newblock {\em Asia-Pacific Financial Markets}, 18:167--189, 2011.
	
	\bibitem{Jovanovskietal2024}
	P.~Jovanovski, A.~Golightly, and U.~Picchini.
	\newblock Towards data-conditional simulation for {ABC} inference in stochastic
	differential equations.
	\newblock {\em Bayesian Analysis}, pages 1--31, 2024.
	
	\bibitem{Kypraios2017}
	T.~Kypraios, P.~Neal, and D.~Prangle.
	\newblock A tutorial introduction to {B}ayesian inference for stochastic
	epidemic models using approximate {B}ayesian computation.
	\newblock {\em Mathematical Biosciences}, 287:42--53, 2017.
	
	\bibitem{lewis1979simulation}
	P.~A.~W. Lewis and G.~S. Shedler.
	\newblock Simulation of nonhomogeneous {P}oisson processes by thinning.
	\newblock {\em Naval Research Logistics Quarterly}, 26(3):403--413, 1979.
	
	\bibitem{Marin2012}
	J.-M. Marin, P.~Pudlo, C.~P. Robert, and R.~Ryder.
	\newblock Approximate {B}ayesian computational methods.
	\newblock {\em Statistics and Computing}, 22(6):1167--1180, 2012.
	
	\bibitem{meddah2024stochastic}
	A.~Meddah.
	\newblock Stochastic hybrid dynamical systems for simulating low-grade glioma
	evolution.
	\newblock {\em PhD thesis}, 2024.
	
	\bibitem{pakdaman2010fluid}
	K.~Pakdaman, M.~Thieullen, and G.~Wainrib.
	\newblock Fluid limit theorems for stochastic hybrid systems with application
	to neuron models.
	\newblock {\em Advances in Applied Probability}, 42(3):761--794, 2010.
	
	\bibitem{PicchiniTamborrino2022}
	U.~Picchini and M.~Tamborrino.
	\newblock Guided sequential {ABC} schemes for intractable {B}ayesian models.
	\newblock {\em Bayesian Analysis}, Advance Publication:1--32, 2024.
	
	\bibitem{R}
	{R Development Core Team}.
	\newblock {\em {R: A Language and Environment for Statistical Computing}}.
	\newblock R Foundation for Statistical Computing, Vienna, Austria, 2011.
	
	\bibitem{Runggaldier2003}
	W.~J. Runggaldier.
	\newblock Chapter 5 - jump-diffusion models.
	\newblock In Svetlozar~T. Rachev, editor, {\em Handbook of Heavy Tailed
		Distributions in Finance}, volume~1 of {\em Handbooks in Finance}, pages
	169--209. North-Holland, Amsterdam, 2003.
	
	\bibitem{SAMSON2025108095}
	A.~Samson, M.~Tamborrino, and I.~Tubikanec.
	\newblock Inference for the stochastic {FitzHugh-Nagumo} model from real action
	potential data via approximate {B}ayesian computation.
	\newblock {\em Computational Statistics \& Data Analysis}, 204:108095, 2025.
	
	\bibitem{singh2010stochastic}
	A.~Singh and J.~P. Hespanha.
	\newblock Stochastic hybrid systems for studying biochemical processes.
	\newblock {\em Philosophical Transactions of the Royal Society A: Mathematical,
		Physical and Engineering Sciences}, 368(1930):4995--5011, 2010.
	
	\bibitem{sisson2018handbook}
	S.~A. Sisson, Y.~Fan, and M.~Beaumont.
	\newblock {\em {Handbook of Approximate {B}ayesian Computation}}.
	\newblock Chapman \& Hall/CRC Handbooks of Modern Statistical Methods. CRC
	Press, Taylor \& Francis Group, 2018.
	
	\bibitem{Sisson2007}
	S.~A. Sisson, Y.~Fan, and M.~M. Tanaka.
	\newblock Sequential monte carlo without likelihoods.
	\newblock {\em Proceedings of the National Academy of Sciences of the USA},
	104(6):1760--1765, 2007.
	
	\bibitem{oksendal2010sde}
	B.~Øksendal.
	\newblock {\em Stochastic Differential Equations: An Introduction with
		Applications}.
	\newblock Springer, Berlin, Heidelberg, 6 edition, 2003.
	
\end{thebibliography}


\newpage
\appendix
\renewcommand{\appendixpagename}{Appendix} 
\appendixpage


\section{Full pseudocode for PDifMP simulation and package information for the SMC-ABC implementation}
\label{app:D:simulationDetails}

This appendix provides explicit pseudocode for the simulation procedures introduced in Section~\ref{sec2:2:Simulation}, together with package details for the implementation of the SMC-ABC algorithm (cf. Section~\ref{sec:implDet}).

\subsection{Pseudocode} \label{appA.1}

\noindent\textbf{SDE path simulation between jumps.} Algorithm \ref{alg:PathFlowPhi} describes the simulation of a discretized SDE path between two consecutive jumps. In particular, it implements the non-uniform time grid defined in Section~\ref{sec2:2:Simulation}, ensuring that the final step size is correctly handled when the jump interval is not a multiple of $h$. This procedure is subsequently called at each iteration of Algorithm \ref{alg:PDifMP_constantLambda} and Algorithm \ref{alg:PDifMP_boundedLambda} (cf. Remark \ref{rem:algorithms}).

\begin{algorithm}[t]
	\caption{SDE: path simulation
	\ \\ \textbf{Input:} Flow $\phi$, jump times $j_k$ and $j_{k+1}$, step size $h$, initial value $x_{j_k}$, constant $z_k$ \\
		\textbf{Output:} Path of SDE between two consecutive jumps 
	}\label{alg:PathFlowPhi}
	\begin{algorithmic}[1]
		\State Set $x[0] = x_{j_k}$
        \State Determine $\tau=j_{k+1}-j_{k}$
            \State Determine $N = \lfloor \tau/h \rfloor$
            \State Determinate $lastStepSize = \tau - N h$
            \For{n=0:(N-1)}
                \State Set $x[n+1] = \phi(h,x[n],z_k)$
            \EndFor
            \State Set $x[N+1] = \phi(lastStepSize,x[N],z_k)$\\
            \Return $x[1:(N+1)]$
	\end{algorithmic}
\end{algorithm}

\begin{remark}\label{rem:algorithms}
    Note that Algorithm \ref{alg:PathFlowPhi} is called iteratively within Algorithms 2 and 3 to generate PDifMP paths over $[0,T]$. Accordingly, the algorithm includes the value $x[N+1] = x_{j_{k+1}}$ in the returned path so that it serves as the initial condition for the subsequent interval. In contrast, the initial value $x[0] = x_{j_k}$ is not returned, since it coincides with the last value of the preceding interval and is therefore already available.
\end{remark}

\noindent\textbf{PDifMP path simulation with constant jump rate.} Algorithm \ref{alg:PDifMP_constantLambda} describes the simulation for a constant jump rate $\Lambda \equiv \lambda$. At each iteration, the next jump time is sampled from an exponential distribution with rate $\lambda$, Algorithm \ref{alg:PathFlowPhi} is called to compute the SDE path, and $z$ is updated according to the transition kernel $Q$. These steps are repeated until the time horizon $T$ is reached.

\begin{algorithm}[htpb]
	\caption{PDifMP with constant jump rate: path and jump time simulation
		\ \\ \textbf{Input:} Flow $\phi$, transition kernel $Q$, jump rate $\lambda$, time horizon $T$, initial values $x_0,z_0$ and step size~$h$ \\
		\textbf{Output:} Path of PDifMP
	}\label{alg:PDifMP_constantLambda}
	\begin{algorithmic}[1]
		\State Set $x[0] = x_0$, $z[0] = z_0$, $j[0] = 0$ and $k = 0$. 
            \State Sample $\tau$ from the exponential distribution with rate $\lambda$. 
            \State Set $j[k+1] = j[k] +\tau$.
            \While{$j[k+1] < T$}
                \State Set $x_{last}$ to the last value of $x$.
                \State Simulate a path of the SDE flow $\phi$ between $j[k]$ and $j[k+1]$ using Algorithm \ref{alg:PathFlowPhi} with step size $h$, initial value $x_{last}$ and constant $z[k]$, and append it to $x$.
                \State Determine $z[k+1]$ from the transition kernel $Q$.
                \State Sample $\tau$ from the exponential distribution with rate $\lambda$. 
                \State Set $k = k + 1$.
                \State Set $j[k+1] = j[k] + \tau$.
            \EndWhile
            \State Set $j[k+1] = T$.
            \State Set $x_{last}$ to the last value of $x$.
            \State Simulate a path of the SDE flow between $j[k]$ and $j[k+1]$ with Algorithm \ref{alg:PathFlowPhi} with step size $h$, initial value $x_{last}$ and constant $z[k]$, and append it to $x$.
            \State Draw $z[k+1]$ from the transition kernel $Q$.\\
            \Return $x$, $z$ and $j$.
	\end{algorithmic}
\end{algorithm}

\medskip

\noindent\textbf{PDifMP path simulation with process-dependent jump rate function.} Algorithm \ref{alg:PDifMP_boundedLambda} extends Algorithm \ref{alg:PDifMP_constantLambda} to a non-constant, bounded jump rate function. Here, candidate jump times are generated using the thinning method, and the algorithm shows explicitly how this rejection procedure is embedded directly into the PDifMP simulation.

\begin{algorithm}[htbp]
	\caption{PDifMP with state-dependent and bounded jump rate function
		\ \\ \textbf{Input:} Flow $\phi$, transition kernel $Q$, jump rate $\Lambda$, upper bound $\lambda$ of $\Lambda$, time horizon $T$, initial values $x_0$, $z_0$ and step size $h$\\
		\textbf{Output:} PDifMP
	}\label{alg:PDifMP_boundedLambda}
	\begin{algorithmic}[1]
		\State Set $x[0] = x_0$, $z[0] = z_0$, $j[0] = 0$, $k = 0$. 
            \State Sample $\tau$ from the exponential distribution with rate $\lambda$.
            \State Set $j_{old} = j[k]$.
            \State Set $j_{new} = j[k] + \tau$.
            \State Set $x_{last}$ to $x_0$
            \While{$j_{new} < T$}
                \State Simulate a path of the SDE flow $\phi$ between $j_{old}$ and $j_{new}$ using Algorithm \ref{alg:PathFlowPhi} with step size $h$, initial value $xlast$ and constant $z[k]$, and append it to $x$.
                \State Set $x_{last}$ to the last value of $x$
                \State Set $accept = \textbf{true}$ with probability $\Lambda(x_{last},z[k])/\lambda$
                \If{accept}
                    \State Draw $z[k+1]$ from the transition kernel.
                    \State Set $j[k+1] = j_{new}$.
                    \State Set $k = k + 1$.
                \EndIf
                \State Sample $\tau$ from the exponential distribution with rate $\lambda$. 
                \State Set $j_{old} = j_{new}$.
                \State Set $j_{new} = j_{old}+ \tau$.
            \EndWhile
            \State Set $x_{last}$ to the last value of $x$
            \State Simulate a path of the SDE flow $\phi$ between $j_{old}$ and $T$ using Algorithm \ref{alg:PathFlowPhi} with step size $h$, initial value $xlast$ and constant $z[k]$, and append it to $x$.
            \State Set $j[k+1] = T$.
            \State Draw $z[k+1]$ from the transition kernel.
            \State Return $x$, $z$ and $j$.
	\end{algorithmic}
\end{algorithm}

\subsection{SMC-ABC package details}\label{appA.2}
The SMC-ABC algorithm is implemented in \texttt{R} \cite{R}, with Algorithms \ref{alg:PathFlowPhi}-\ref{alg:PDifMP_boundedLambda} coded in \texttt{C++}, using the package \textrm{Rcpp} \cite{Rcpp}. Parallelisation is achieved using the \texttt{foreach} and \texttt{doParallel} packages, and the multivariate Gaussian perturbation kernel is implemented using the \texttt{mvnfast} package. Following \cite{Buckwar2020}, estimates of the invariant density and spectral density are obtained using the \texttt{R} functions \texttt{density} and \texttt{spectrum}, respectively.


\section{Robustness under misspecification of observed jump count}
\label{app:A:Robustness}

\begin{figure}[t]
    \centering
    \includegraphics[width=1.0\textwidth]{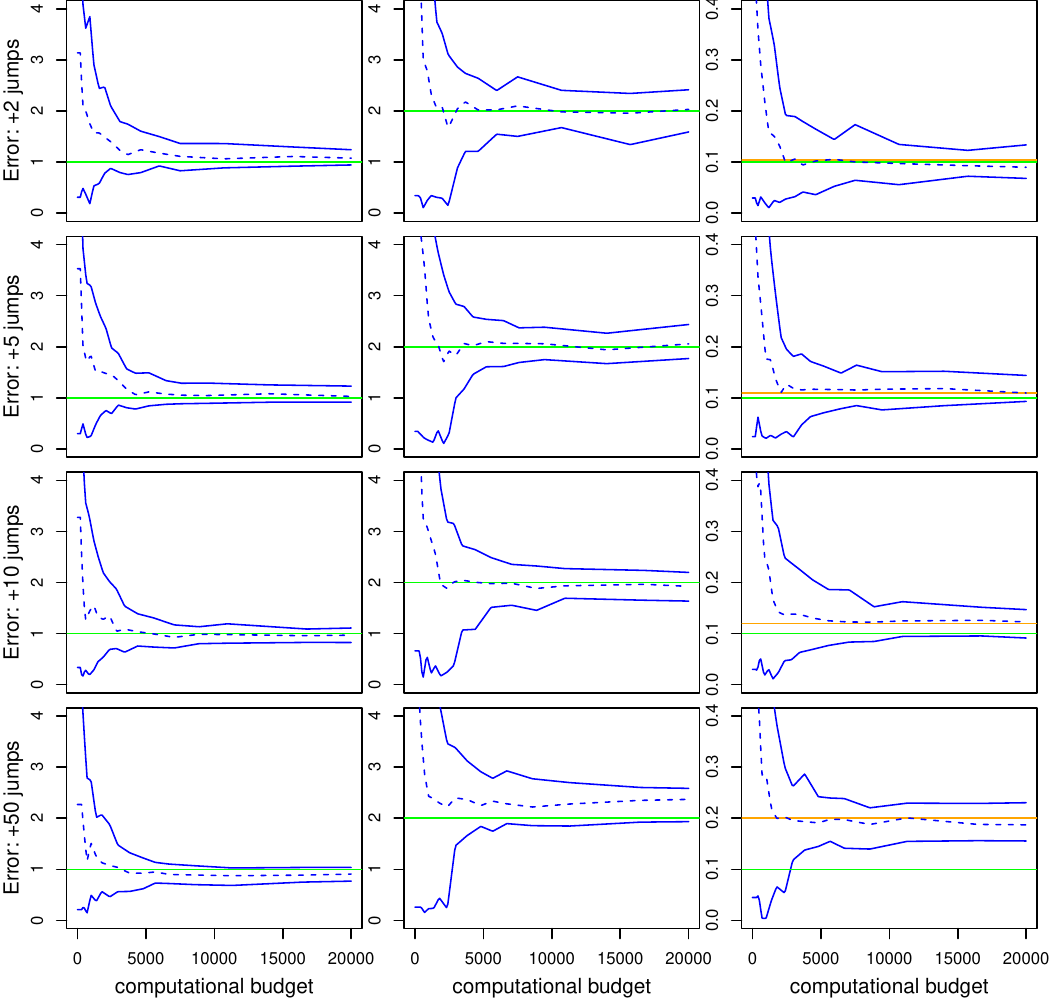}
    \caption{90\% CIs as functions of the computational budget of the marginal posterior densities of $\theta$ \eqref{eq:theta} for TP~1 (OU-PDifMP) with misspecified number of jumps. }
    \label{fig:CI_noise}
\end{figure}

To assess the sensitivity of the inference results to misspecification of the observed number of jumps, we repeat the inference for TP1 (OU-PDifMP) using different misspecified values of $N^j$ in~\eqref{eq:observed_y}. Figure~\ref{fig:CI_noise} shows the resulting 90\% credible intervals when the observed jump count differs from the true count by 2, 5, 10, and 50 jumps, respectively. The last case represents a substantial misspecification, corresponding to approximately twice the actual number of jumps.

Across the considered cases, the credible intervals for $\sigma$ and $b$ continue to cover the true parameter values (green lines), and the posterior distributions remain closely centred around these values, with only a minor bias appearing under the most pronounced misspecification. In contrast, the credible interval for the jump-rate parameter $\lambda$ shifts systematically with the misspecified jump count, as expected from the direct dependence of $\lambda$ on the number of observed jumps. To illustrate this dependence, the orange line indicates the jump rate corresponding to each misspecified jump count. The posterior estimates of $\lambda$ shift towards these values, while the inference for $\sigma$ and $b$ remains robust to the misspecification of $N^j$.


\section{Variability across multiple independent runs}
\label{app:C:MulipleRuns}

\begin{table}
    \centering
    \begin{tabular}{|llcccc|}
    \hline
    Test problem & Parameter & True Value & Mean $\pm$ SD & 5\% & 95\% \\
    \hline
    \textbf{TP1} (OU-PDifMP) & $\sigma$ & 1 & $1.053 \pm 0.104$ & 0.846 & 1.254 \\
        & $b$ & 2 & $1.978 \pm 0.106$ & 1.617 & 2.319 \\
        & $\lambda$ & 0.1 & $0.097 \pm 0.020$ & 0.062 & 0.147 \\
    \hline
    \textbf{TP2} (WDSHO-PDifMP) & $\sigma$ & 1 & $0.925 \pm 0.058$ & 0.753 & 1.134 \\
        & $b$ & 10 & $9.767 \pm 0.312$ & 6.492 & 13.396 \\
        & $\lambda$ & 0.1 & $0.102 \pm 0.006$ & 0.074 & 0.126 \\
    \hline
    \textbf{TP3} (WPWD-PDifMP) & $\sigma$ & 1 & $0.979 \pm 0.023$ & 0.767 & 1.140 \\
    & $b$ & 2 & $2.022 \pm 0.029$ & 1.946 & 2.087 \\
    & $\lambda$ & 0.1 & $0.094 \pm 0.007$ & 0.079 & 0.111 \\
    \hline
    \textbf{TP4} (SwitchedSHO-PDifMP) & $\sigma$ & 1 & $1.098 \pm 0.199$ & 0.768 & 1.402 \\
        & $b$ & 0.1 & $0.128 \pm 0.051$ & 0.056 & 0.229 \\
        & $\lambda$ & 0.1 & $0.106 \pm 0.005$ & 0.077 & 0.137 \\
    \hline
\end{tabular} \\
    \caption{Posterior statistics for the parameters $\sigma$, $b$, and $\lambda$ of TP1--TP4 across five independent runs. Mean $\pm$ SD denotes the mean and standard deviation of the posterior means across runs, while the 5\% and 95\% columns report the averaged posterior quantiles across runs.}
    \label{tab:tableOfVariability}
\end{table}

In Section~\ref{sec:5:Inference}, only a single representative run of the SMC-ABC algorithm was shown for visual clarity. To assess the variability of the results across independent runs, Table~\ref{tab:tableOfVariability} summarizes the ABC posterior estimates obtained from five independent runs for all four test problems. The reported mean values are the averages of the posterior means across the five runs, while the $\pm$SD values denote their standard deviations and thus quantify the inter-run variability. The 5\% and 95\% columns report the corresponding averaged bounds of the 90\% credible intervals. Note that for TP3 we used the observed data \eqref{eq:observed_y_extended} and the TP3-specific summary set \eqref{eq:summaries_extended}.

Overall, the posterior means remain close to the true values and the estimates show consistently low inter-run variability across the five runs for all four test problems. Importantly, the averaged 90\% credible intervals contain the true parameter values for all parameters and test~problems.


\section{Impact of the time horizon}
\label{app:B:timeHorizon}

\begin{figure}[t]
    \centering
    \includegraphics[width=1.0\textwidth]{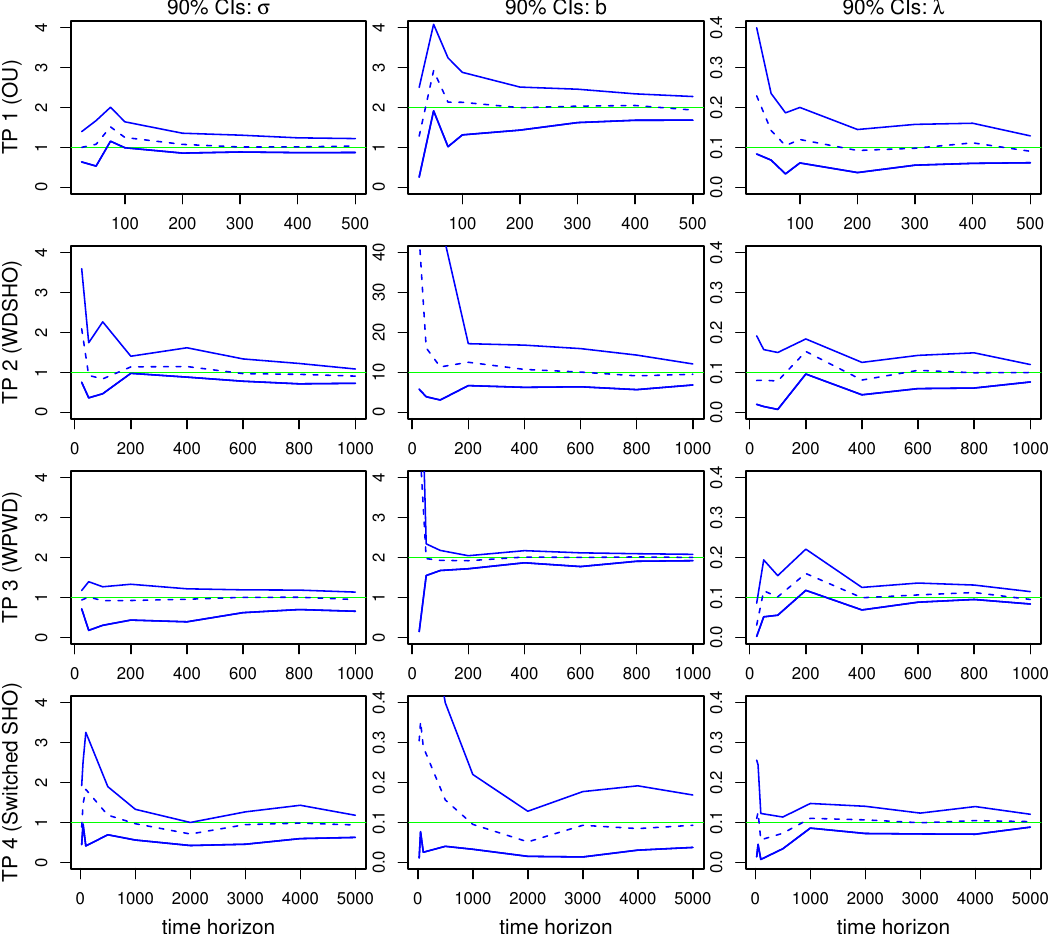}
    \caption{90\% CIs as functions of the time horizon of the marginal posterior densities of $\theta$ \eqref{eq:theta} for the four test problems with fixed computational budget. }
    \label{fig:CI_T}
\end{figure}

To investigate the impact of the time horizon on the inference results under a fixed computational budget, we compare the 90\% credible intervals obtained for increasing observation horizons. Figure~\ref{fig:CI_T} shows that, across the considered test problems, increasing the time horizon generally leads to narrower credible intervals and posterior estimates that are better centred around the true parameter values. The extent of this improvement varies across parameters and test problems. Note that for TP3 we used the observed data \eqref{eq:observed_y_extended} and the TP3-specific summary set \eqref{eq:summaries_extended}.

\end{document}